\documentclass[prb,aps,twocolumn,home,floats]{revtex4}

\usepackage{graphics}
\begin{document}

\title{{\rm\small\hfill to appear in {\em Nanocatalysis: Principles, Methods, Case Studies},}\\
{\rm\small\hfill U. Heiz, H. Hakkinen and U. Landman (Eds.),}\\
{\rm\small\hfill Springer, Berlin}\\
\quad \\
Nanometer and sub-nanometer thin oxide films\\ at 
       surfaces of late transition metals}

\author{Karsten Reuter}

\affiliation{Fritz-Haber-Institut der Max-Planck-Gesellschaft,\\ 
Faradayweg 4-6, D-14195 Berlin-Dahlem (Germany)}

\received{19 September 2004}

\begin{abstract}
If metal surfaces are exposed to sufficiently oxygen-rich environments, oxides start to form. Important steps in this process are the dissociative adsorption of oxygen at the surface, the incorporation of O atoms into the surface, the formation of a thin oxidic overlayer and the growth of the once formed oxide film. For the oxidation process of late transition metals (TM), recent experimental and theoretical studies provided a quite intriguing atomic-scale insight: The formed initial oxidic overlayers are not merely few atomic-layer thin versions of the known bulk oxides, but can exhibit structural and electronic properties that are quite distinct to the surfaces of both the corresponding bulk metals and bulk oxides. If such nanometer or even sub-nanometer surface oxide films are stabilized in applications, new functionalities not scalable from the known bulk materials could correspondingly arise. This can be particularly important for oxidation catalysis, where technologically relevant gas phase conditions are typically quite oxygen-rich. In such environments surface oxides may even form naturally in the induction period, and actuate then the reactive steady-state behavior that has traditionally been ascribed to the metal substrates. Corresponding aspects are reviewed by focusing on recent progress in the modeling and understanding of the oxidation behavior of late TMs, using particularly the late $4d$ series from Ru to Ag to discuss trends.
\end{abstract}

\maketitle

\section{Introduction} 

Confinement to nanoscale dimensions and the resulting quantization of electrons gives rise to two-dimensional quantum films, one-dimensional quantum wires, or zero-dimensional quantum dots. All three situations may exhibit properties and functionalities that make them different from the corresponding bulk material. 
In this respect, research on nanocatalysis is inspired by the expectation that due to such a reduced dimensionality of the active material new catalytic behavior can arise, which is not scalable from bulk-like properties. \cite{haruta97} A salient target of such investigations are nanometer-size clusters of atoms, representing the three-dimensional confinement situation. While the potential of such clusters for catalytic applications is just being explored, it has already been recognized that they often exhibit the undesirable propensity to coarsen into larger aggregates on the support material. This problem could be less severe for the case of confinement in only one dimension, i.e. when striving for appealing properties of nanometer or sub-nanometer thin films. In fact, at the catalyst surface such films may even form automatically during the induction period, as a consequence of the exposure to the operating conditions. A prominent example of this is oxidation catalysis, where the oxygen-rich environment might oxidize the sample. Recent experimental and theoretical studies noted that particularly at late transition metals (TMs) few or even one atomic-layer thin, so-called surface oxide films are formed, when these are held in a realistic gas-phase environment under catalytic or alike $(T,p)$-conditions. The vertical confinement makes these films distinct to surfaces of both bulk metals and bulk oxides, and their specific properties can have beneficial, as well as detrimental effects on the desired functionality.

Appropriately tailoring and exploiting these functionalities requires an 
atomic-scale characterization and understanding of the oxidation process, as well as of the resulting surface composition and stoichiometry (under the steady-state reaction conditions). Although oxide formation has been of technological and scientific interest since long, such an understanding is, however, only just emerging, and the underlying microscopic processes are unfortunately still largely unclear for everything that goes beyond 
sub-monolayer oxygen adsorbate phases on low-index single crystal surfaces. This holds in particular for the transition from an oxygen adsorbate layer to the surface oxide. A main reason behind this lack of understanding is that 
atomic-scale investigations of TM surfaces on the verge of oxide formation still pose a significant challenge for surface scientists: The need for rather high oxygen partial pressures (high compared to those possible in standard surface science experiments) and elevated temperatures required to initiate the process, a low degree of order at the surface, and often complex, large unit-cell geometries name but a few of the obstacles encountered. Since this often exceeds the analytical capabilities of one experimental technique alone, multi-method approaches (e.g. combining scanning, diffraction and spectroscopic measurements) have become an important tool in this field. In addition, first-principles calculations, almost exclusively employing density-functional theory (DFT), have played a key role in this research area, in particular when suitably combined with concepts from thermodynamics or statistical mechanics. Although most of the arguments and understanding presented in this review are based on such calculations, I will not describe the methods themselves here, but refer to excellent monographs (see for example Refs. \onlinecite{dreizler90} and \onlinecite{parr94}) for DFT itself, to Refs. \onlinecite{scheffler00} and \onlinecite{gross02} for applications to surfaces, and to Ref. \onlinecite{reuter04a} for the hybrid statistical mechanics methods.

Instead, this paper will focus on recent progress in the modeling and understanding of oxide formation of late TMs, using particularly the late $4d$ series from Ru to Ag to discuss trends. Section II will highlight aspects leading to the formation of surface oxides, especially on-surface adsorption and accommodation of oxygen in the sub-surface region. Key factors (also for the surface oxide stability itself) are the bulk-oxide heat of formation, diffusion energy barriers, elastic strain energies, and the nature of the surface-oxide--metal interface. Sometimes the nano- or sub-nanometer thickness of the formed oxide film is ruled by the thermodynamic conditions, but often the thickness is determined by kinetics, e.g. by the diffusion of metal or oxygen atoms (between interface and surface), or by the on-going oxidation reaction at the surface in a catalytic environment. Examples for both situations will be discussed in Section III, where we will also see that such thin oxide films can have a different structure to what is known from the stable bulk oxides. Setting these results into a ``catalytic perspective'', I will conclude in Section IV, for example that the catalytic activity of Ru in high-pressure oxidation reactions is likely due to nanometer thin, but nevertheless bulk-like oxide patches, whereas for Pd and Ag the catalysis is more likely actuated by novel, surface-confined sub-nanometer thin oxidic structures.

\section{Initial oxidation of transition-metal surfaces}

\subsection{Formation of adlayers}

\begin{figure}
\scalebox{0.22}{\includegraphics{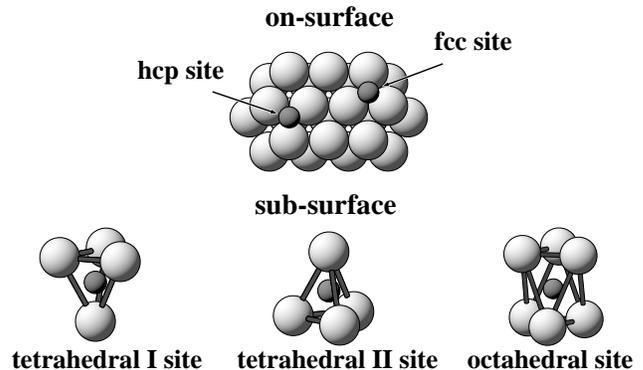}}
\caption{Highly-coordinated adatom sites at a fcc(111) or hcp(0001) surface. 
The upper picture shows a top view of the surface, labeling the two threefold coordinated hollow sites. The lower row illustrates the local geometries of the three high-symmetry interstitial sites between the first and second substrate layer. Metal atoms are depicted as light, large spheres, and oxygen atoms as dark, small spheres.}
\label{sites}
\end{figure}

Oxide formation at metallic surfaces begins with the dissociation of the O$_2$ molecule. This process is still not fully understood (see for example Ref. \onlinecite{behler-2004} and references therein), but for the present review we don't need to discuss this in detail. As the end product of the dissociation atomic oxygen is chemisorbed at highly coordinated surface sites. I will mainly address the lowest energy surfaces of the late $4d$ TMs which have the hcp(0001) geometry for Ru, and the fcc(111) geometry for Rh, Pd, and Ag. For these systems the oxygen adatoms occupy threefold hollow sites which are sketched in Fig. \ref{sites}. These two adsorbate positions, named hcp and fcc sites, differ only with respect to the second substrate layer: Below the hcp hollow site there is a second-layer atom, and below the fcc hollow site there is none. In other words, these sites represent positions atoms would occupy, if the crystal were continued in a hcp or a fcc stacking sequence. Interestingly, on the four $4d$ materials covered in this paper, oxygen prefers initially those threefold hollow sites that correspond to the stacking type of the substrate, i.e., the hcp hollow sites on Ru and the fcc hollow sites on Rh, Pd, and Ag. With increasing number of oxygen adatoms repulsive lateral interactions between the adsorbates lead to the formation of ordered superstructures, typically with a $(2 \times 2)$ surface unit-cell at low coverages. Then, depending on the material and whether higher coverage on-surface structures are formed at all, this may be followed by either $(2 \times 1)$ or $(\sqrt{3} \times \sqrt{3})R30^\circ$ periodic overlayers, with oxygen atoms still occupying either hcp or fcc hollow sites.

The quantity ruling the adsorption-site preference and the formation of ordered adlayers is the adatom binding energy:
\begin{equation}
E_b(\Theta) = - \frac{1}{N_{\rm O}} \left[ E_{\rm O@surf.} - E_{\rm surf.} - 
(N_{\rm O}/2) E_{\rm O_2} \right]  .
\label{bindeng}
\end{equation}
Here, $N_{\rm O}$ is the total number of O atoms in the unit-cell at the considered coverage $\Theta$ (measured in monolayers, ML, i.e. the fraction compared to the number of first layer metal atoms). $E_{\rm O@surf.}$, $E_{\rm surf.}$, and $E_{\rm O_2}$ are the total energies of the adsorbate system, of the corresponding clean surface before adsorption, and of the isolated oxygen molecule. A positive binding energy $E_b(\Theta)$ reflects that the dissociative adsorption of O${}_2$ is exothermic. Equation (\ref{bindeng}) can be evaluated by DFT, and for big systems this is in fact the only approach that is currently feasible and predictive. One has to stress, however, that for oxygen at transition metal surfaces such calculations are still demanding. Without describing the technical challenges of performing such large-scale DFT studies, I only mention the basic problem: Though DFT is in principle exact, in practice it is not, because the exchange-correlation (xc) functional is (and probably ever will be) only known approximately. For many systems this is not a severe problem, but for the O$_2$ molecule even present-day gradient-corrected (GGA) functionals suffer from a binding energy error of about 0.5\,eV per O atom. \cite{perdew-PRL-1996,zhang-PRL-1998,perdew-PRL-1999} Though some error cancellation will occur in eq. (\ref{bindeng}), this O$_2$ problem calls for caution when judging on the exo- or endothermicity of oxygen adsorption or oxide formation. Binding energy {\em differences} between different adsorbate systems, for which the gas phase O$_2$ problem cancels out at least partially, appear fortunately often much less affected by the xc approximation. Still, caution is advisable, and if doubts exist, a regional xc correction, e.g., following the approach of Filippi {\em et al.} \cite{filippi-PRL-2002}, may be necessary.

\begin{figure}
\scalebox{0.67}{\includegraphics{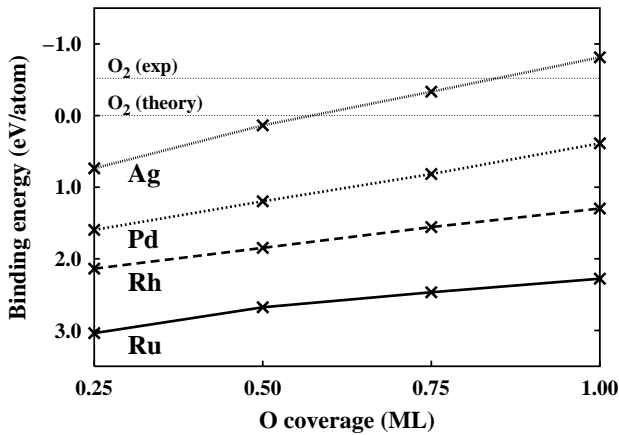}}
\caption{Computed (DFT-GGA) binding energies $E_b(\Theta)$ for on-surface oxygen
chemisorption into the most stable hollow sites of the basal surface of the late $4d$ TMs (Ru = hcp site, Rh/Pd/Ag = fcc site) and employing $(2 \times 2)$ surface unit-cells (the lines are only to guide the eye). The energies are given with respect to the theoretical O${}_2$ binding energy, cf. eq. (\ref{bindeng}), while the upper horizontal dotted line indicates the zero-level, if the experimental O${}_2$ binding energy were used (after Ref. \onlinecite{todorova02}).}
\label{miraprl1}
\end{figure}

Evaluating eq. (\ref{bindeng}) is most useful for identifying adsorption sites, lateral interactions or the saturation coverage, and by such calculations the interplay between theory and experiment has indeed been most stimulating and synergetic in this field. The atomic scale characterization of higher coverage O adsorbate layers forms already an intriguing example for this. Although not really fully addressed and understood, it seems that the sticking coefficient for O$_2$ drops either dramatically after a certain threshold coverage at the surface is reached (Ru, Rh, Pd) or is very low from the beginning (Ag). Particularly the sharp drop after reaching a coverage of about 0.5\,ML at Ru(0001) and Rh(111) leads to an apparent uptake saturation, when employing low gas exposures as typical for many early ultra-high vacuum (UHV) surface science studies.\cite{surnev85,pfnur89,comelli98} DFT calculations predicted, however, that much higher coverages should still be possible 
\cite{stampfl-PRB-1996,stampfl-PRL-1999,ganduglia99}, and Fig. \ref{miraprl1} shows such data for some selected adsorbate configurations spanning the whole coverage range up to 1\,ML \cite{todorova02}. Although the decreasing binding energies with coverage indicate repulsive interactions between the adsorbed O atoms, they are still clearly exothermic up to the full ML for Ru (and also for Rh). As for the O/Ru case adsorption into other sites (e.g. below the surface) or other structures can be ruled out (see below), these results imply that energetically the formation of denser overlayers is possible, well beyond the apparent saturation at $\Theta \approx 0.5$\,ML. And in fact, by using more oxidizing gases that create higher partial pressures (e.g. NO$_2$), orders of magnitude higher O$_2$ exposures (sometimes millions of Langmuirs) or the more novel, direct {\em in-situ} studies at elevated pressures such overlayers (and configurations corresponding to even much higher O uptakes) are now routinely observed experimentally. One should stress, that on all four TM surfaces discussed in this paper, such experiments have opened a new door to study the oxidation beyond sub-monolayer overlayers in a controlled fashion, significantly improving the scientific understanding over the last years.

\begin{figure*}
\scalebox{0.45}{\includegraphics{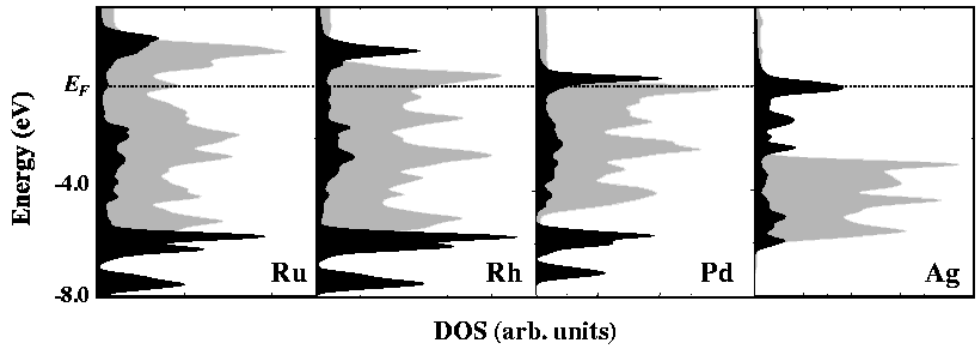}}
\caption{Computed (DFT-GGA) projected density of states (PDOS), cf. eq. (\ref{eq:DOS_a}), of the clean basal surfaces of the late $4d$ TMs (gray shading), and of the O$(1 \times 1)$ adlayer (black shading). Shown are the $N_{\alpha}(\epsilon)$ contributions of the Ru $4d$ and O $2p$ states. The decrease of the oxygen PDOS band width (from left to right) and the increased filling of antibonding states (the higher-energy oxygen PDOS region) is clearly visible.}
\label{tmdos}
\end{figure*}

Figure \ref{miraprl1} also illustrates the mentioned O$_2$-molecule problem. Using in eq. (\ref{bindeng}) the {\em experimental} O$_2$ total energy, $E_{\rm O_2}$, instead of the theoretical one, would shift the energy zero to the upper dotted horizontal line. This is a too extreme alternative though, because some xc error also exists in $E_{\rm O@surf.}$ and $E_{\rm surf.}$. Thus, the true energy zero is probably somewhere in between the two dotted lines. Obviously, in particular for silver the inaccuracy in the molecular binding requires special attention, even when only discussing the sign of the binding energy \cite{li02}. What is, however, unambiguously evident from Fig. \ref{miraprl1} is that the oxygen-metal bond strength decreases steadily from Ru to Ag, and this is in fact known qualitatively since long. The reason behind it is clearly visible in the projected density of states (DOS), which is defined as \cite{scheffler00}
\begin{equation}
N_{\alpha}(\epsilon) = \sum_{i= 1}^{\infty} |\langle \phi_\alpha | \varphi_i \rangle|^2 \delta(\epsilon -\epsilon_i)   .
\label{eq:DOS_a}
\end{equation}
Here $\varphi_i$ are the Kohn-Sham orbitals, and $\phi_\alpha$ is a properly chosen localized function, in the present case most suitably the Ru $4d$ and the O $2p$ atomic orbitals within a sphere around given atomic sites. According to the Anderson-Grimley-Newns model of chemisorption the interaction of the O $2p$ level with the narrow $4d$ band of the TM surface gives rise to bonding states at the lower edge of the $d$-band and antibonding states at the upper edge of the $d$ band. The corresponding peaks can be readily identified in Fig. \ref{tmdos}, illustrating the projected DOS of a high coverage O adsorbate layer for the late $4d$ TM series. The Fermi level moves from the $d$-band center (for Ru it is already at a $\approx 70$\% filling) to about 3.5 eV above the top of the $d$ band (for the noble metal Ag), and thus the antibonding oxygen--metal states become progressively more and more occupied. \cite{scheffler00,norskov90,hammer01} As a consequence, the bond strength is weakened, explaining the steadily decreasing binding energies of Fig. \ref{miraprl1}. For the discussion below, it is finally worthwhile mentioning that despite these large variations of the binding energy with element (as well as with coverage), the oxygen-metal bond length does not change appreciably and stays always roughly at 2\,{\AA}.

\subsection{Oxygen accommodation below the top metal layer}

\begin{figure*}
\scalebox{0.4}{\includegraphics{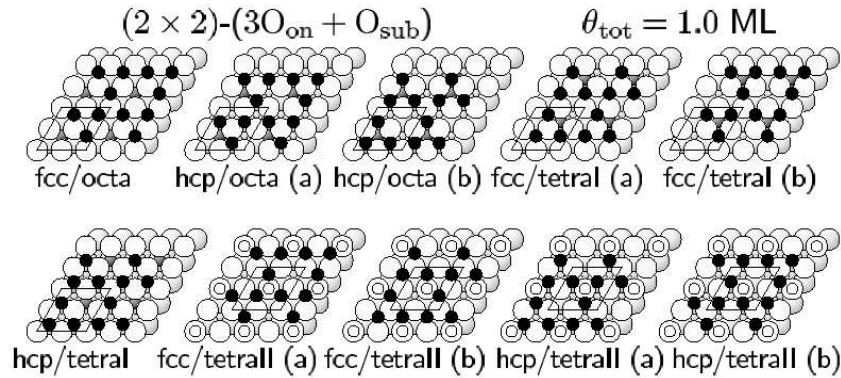}}
\caption{Top view of all possible on-surface/sub-surface site combinations 
at a total coverage $\Theta = 1$\,ML with one oxygen atom per $(2 \times 2)$ surface unit-cell located below the surface, i.e. corresponding to a fraction of 25\,\% below the surface (see Fig. \ref{sites} for the explanation of the different sites). The metal atoms are drawn as big spheres (white for the surface layer, gray for second layer), and the oxygen atoms are drawn as small spheres (black for on-surface, gray for sub-surface). Oxygen atoms in tetra-II sites below the first layer metal atoms are invisible in this plot and are schematically indicated by small white circles. Symmetry inequivalent occupation of the same kind of on- and sub-surface sites are denoted with (a) and (b), respectively.}
\label{verosites}
\end{figure*}

It is often assumed that deposition of oxygen on metal surfaces gives initially rise to stable overlayers (either ordered or disordered). Apparently this is often correct, but there is no reason that it holds in general. In fact, oxygen atoms may well go below the surface, either staying in the near-surface region or dissolving into the bulk. Or, metal-oxide patches could form immediately.
Experimentally, very little is unfortunately still known about this. For the basal surfaces of the late $4d$ transition metals recent DFT-GGA calculations indicate that initially oxygen adatoms stay at the surface, and only when the coverage exceeds a certain value some adatoms will go sub-surface.\cite{todorova02} They then remain preferentially close to the surface, i.e., they will not dissolve into the bulk.\cite{reuter02a,li03}  The interaction between these sub-surface oxygen atoms is attractive, and this situation may well already be called a surface oxide, or ``stage I'' of oxide formation. 

In principle, oxygen may occupy interstitial or substitutional sites in the crystal, but mostly due to the high cohesive energy of the late TMs the latter are found to be energetically much less favorable.\cite{li02} Thus, they will not play a role in thermodynamic equilibrium, or close to it, and we may focus our discussion on interstitial sites. Directly below the topmost metal layer, i.e. in the immediate sub-surface region, there are three high-symmetry possibilities for such interstitial sites (cf. Fig. \ref{sites}). Namely, there is an octahedral site (henceforth called octa) and a tetrahedral site (tetra-I) directly below the on-surface fcc and hcp site, respectively. And there is a second tetrahedral site (tetra-II) directly below a first layer metal atom. The main trends in the energetics of on- versus sub-surface oxygen are already discernible by comparing surface structures containing from one up to four O atoms in a model subset of $(2 \times 2)$ unit-cells. Of these up to four adatoms, consider that any one can be located either on- or below the surface in any of the high-symmetry sites displayed in Fig. \ref{sites}. Despite the ostensible simplicity of this model subset, the number of possible combinations of on- and sub-surface oxygen atoms in various site and symmetry configurations is still quite large, as exemplified in Fig. \ref{verosites} by all configurations where one out of a total of four oxygen atoms is located below the surface. Interestingly, from all these structural possibilities the combination of on-surface O in fcc sites and sub-surface O in tetra-I sites was in all four late $4d$ metals found to be energetically either most stable or very close to the most stable geometry.\cite{todorova02,reuter02a,li02,ganduglia02,todorova04} 

\begin{figure}
\scalebox{0.21}{\includegraphics{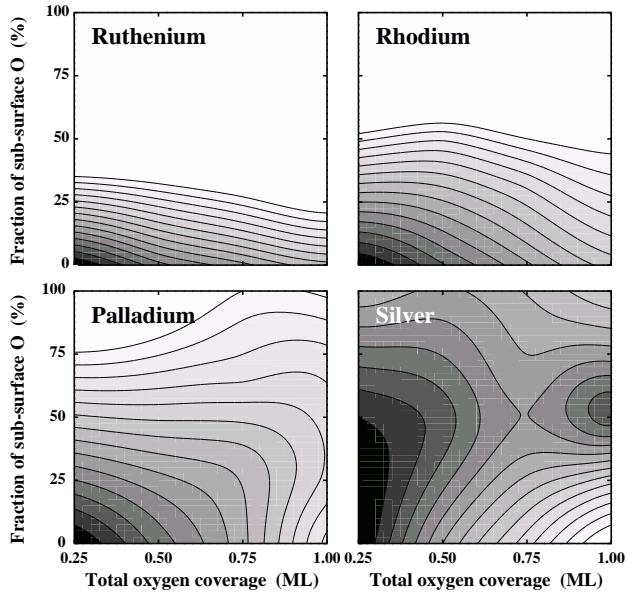}}
\caption{Averaged binding energy $E_b$ (using DFT-GGA and eq. (\ref{bindeng}))
as a function of the total O coverage $\Theta$ of which a certain percentage is located below the basal surface (on-surface in fcc and sub-surface in tetra-I sites). The highest $E_b$ are always found for the pure on-surface chemisorption phase at $\Theta = 0.25$\,ML (chosen as zero reference, black area), with each contour line (lighter gray areas) at 0.1\,eV steps towards less stable $E_b$ (from Ref. \onlinecite{todorova02}).}
\label{miraprl2}
\end{figure}

Occupation of sub-surface sites will commence at a certain total coverage $\Theta_c$, when a structure with a non-zero fraction of the O atoms below the surface is energetically more favorable than the pure on-surface adsorption (corresponding to a zero fraction of oxygen atoms below the surface). Figure \ref{miraprl2} summarizes the energetics within the described model subset at the basal surface, and we see that for all four late $4d$ TMs on-surface adsorption is initially more favorable than sub-surface incorporation. However, with increasing coverage this picture changes as the aforedescribed repulsive lateral interactions between the on-surface oxygen atoms progressively diminish the preference for adsorption into these sites. As can be deduced from Fig. \ref{miraprl2} this threshold coverage $\Theta_c$, is beyond 1\,ML for Ru and Rh, i.e. oxygen incorporation will only start after completion of a full O$(1 \times 1)$ overlayer, in nice agreement with experimental findings.\cite{gibson99,stampfl-PRL-1996} Close to $\Theta = 1$\,ML the energy differences are very small in the case of Rh though. Hence, a finite concentration of sub-surface oxygen might already be present at total coverages even slightly below 1\,ML at elevated temperatures.\cite{ganduglia02,wider98}
For Pd, $\Theta_c$ lies around 3/4\,ML, and for Ag sub-surface incorporation starts almost immediately (cf. Fig. \ref{miraprl2}).

\begin{figure}
\scalebox{0.30}{\includegraphics{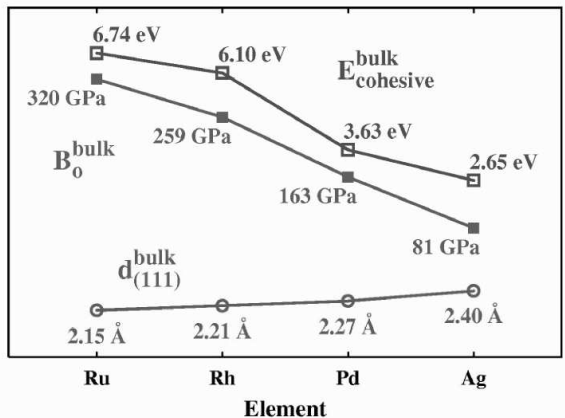}}
\caption{Trend of DFT-GGA computed bulk modulus, cohesive energy,
and layer distance in (111) direction for the late $4d$ TMs.}
\label{bulkmod}
\end{figure}

This trend for sub-surface accommodation of oxygen atoms at the late $4d$ TMs can be understood by looking at the sub-surface geometries. In all relaxed structures the sub-surface oxygen-metal bond lengths are slightly larger than 2\,{\AA}, i.e. they are similar to the on-surface O--metal distances. Comparing these bond lengths with the available geometrical space inside a frozen metal lattice, one finds them to be rather incompatible. For the unrelaxed tetra-I site of the late $4d$ metals the bond length would range from 1.65 {\AA} (Ru) to 1.80 {\AA} (Ag). Thus, oxygen incorporation necessarily must go together with a substantial local expansion of the metal lattice. As shown by Todorova {\em et al.} \cite{todorova02} the bond strength of sub-surface oxygen atoms may therefore roughly be divided into a stabilizing contribution due to the chemical binding and a destabilizing contribution from the lattice deformation cost. Going from Ru to Ag the first component should follow the same $d$-band-filling trend of the oxygen-metal bond strength that was discussed in Section IIA for on-surface oxygen. And also the latter component follows this trend, i.e., it scales roughly as the bulk cohesive energy or the bulk modulus, which are largest for Ru and smallest for Ag as noted in Fig. \ref{bulkmod}. The two contributions (O--metal bond strength and metal-lattice distortion energy) have opposite sign and apparently roughly compensate each other, leading to a rather similar energetic stability of sub-surface oxygen in all four elements.\cite{todorova02}

Due to the lattice-deformation cost sub-surface oxygen turns out initially less favorable than on-surface adsorption, but with increasing coverage the repulsive lateral interactions for on-surface oxygen drive this preference down until eventually penetration of some oxygen adatoms becomes favorable. As the on-surface bond strength decreases from Ru to Ag, while the sub-surface bond strength stays roughly constant, this crossover will occur progressively earlier, i.e. $\Theta_c$ is large for Ru and small for Ag. 

If oxygen penetrates, the deformation cost also rationalizes, why sub-surface incorporation was found to be more favorable compared to bulk dissolution. Although the latter occurs of course always to some extent just on entropy grounds, there will at any rate be an enrichment of oxygen atoms in the easier to relax interstitial sites in the immediate near-surface fringe. And one could expect this deformation argument to hold even more for sub-surface sites close to even lower coordinated atoms, i.e. interstitials in the vicinity of point defects, steps or dislocations. The latter are in fact frequently believed to be {\em the} nucleation centers for surface oxide formation, and kinetic arguments like an easier O penetration are often put forward as explanation. Yet, we see that the thermodynamic deformation cost factor could also favor an initial oxygen accommodation close to such sites.

\subsection{Oxygen accumulation in the surface region and surface oxide formation}

Thermodynamically the bulk oxide will start to form at the oxygen coverage, $\Theta_c^{\rm thd}$, at which the average oxygen binding energy, cf. eq. (\ref{bindeng}), equals the oxide heat of formation \cite{carlisle00a}. The word ``coverage'' is used here in a wider sense, i.e. meaning the total amount of oxygen atoms in the near-surface region. Interestingly, the just discussed $\Theta_c$ for the initial incorporation of oxygen at sub-surface sites in the late $4d$ metals does not only follow $\Theta_c^{\rm thd}$ in trend, but also in approximate value \cite{todorova02}. This suggests sub-surface oxygen as an intermediate, or a precursor, for the formation of the bulk oxide or other oxidic structures beyond on-surface adsorbate layers. Recent experimental studies seem to support this view, i.e., they indicate that thin oxidic structures, now mostly coined surface oxides, start to form roughly at coverages $\Theta_c$.\cite{boettcher97,over00,gustafson04,lundgren02,carlisle00b}

\begin{figure}
\scalebox{0.55}{\includegraphics{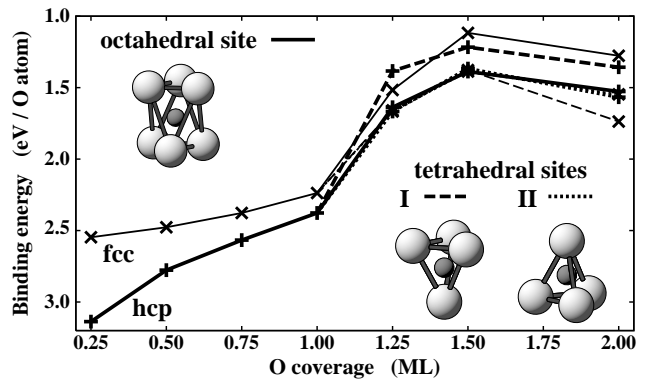}}
\caption{Binding energies, calculated by DFT-GGA, for oxygen at
Ru(0001). Coverages with $\Theta \le 1$\,ML correspond to pure on-surface adsorption (hcp or fcc site). For $\Theta > 1$\,ML an O$(1 \times 1)$ arrangement is always present at the surface, cf. Fig. \ref{miraprl2}, while the remaining O is located in either of the three interstitial sites between the first and second substrate layer. From the six possible structures at each coverage, the three with on-surface O in hcp (fcc) sites are drawn with thicker (thinner) lines (from Ref. \onlinecite{reuter02a}).}
\label{subORu}
\end{figure}

For the ``true'' bulk oxide a higher oxygen concentration might be required, but for oxide nucleation and for the formation of oxidic islands such increased oxygen content is only needed locally. In this respect it is remarkable that although for the on-surface adsorption at the late $4d$ TM basal surface the adatom-adatom interaction is repulsive (see the discussion of Fig. \ref{miraprl1} above), an attractive interaction was found for sub-surface oxygen atoms in Ru(0001) and Rh(111) \cite{reuter02a,ganduglia02}. This is illustrated in Fig. \ref{subORu}, which shows the course of the average binding energy of O at Ru(0001) upon increasing the amount of oxygen atoms between the first and second substrate layer up to 1\,ML sub-surface coverage, i.e., 2 ML total coverage. As discussed in the last section, for Ru(0001) oxygen penetration starts only after the full on-surface monolayer coverage is reached. Up to $\Theta = 1$ML the binding energy decreases steadily due to the repulsive lateral interactions. For even higher coverages, i.e. upon the ensuing filling of sub-surface sites, the binding energy curve exhibits an inflection. This implies overall attractive interactions that favor an accumulation of sub-surface oxygen in dense two-dimensional islands with a local $(1 \times 1)$ periodicity.\cite{reuter02a}

While for Ru(0001) \cite{reuter02a} and Rh(111) \cite{ganduglia02} the attractive interactions lead to dense oxygen islands with a local $(1 \times 1)$ ordering below the surface, cf. Fig. \ref{subORu}, at Ag(111) it already becomes more favorable to incorporate oxygen rather between deeper layers, when a local sub-surface oxygen coverage of $\sim 1/4$\,ML is exceeded.\cite{li03} Apparently, each metal is only able to sustain a certain amount of oxygen per layer, and the differences between Ru and Ag are significant. Whereas Ru(0001) can accommodate a full oxygen monolayer on the surface plus a full oxygen monolayer below the top Ru layer, for Ag(111) this is only 25\% of a monolayer on and 25\% below the top Ag layer. What seems to be a more general feature, however, is the O-metal-O trilayer type structure, that results from the accumulation of sub-surface oxygen below the topmost, oxygen-adsorbate covered metal layer. This accumulation will eventually lead to the formation of an ordered structure, which as already mentioned above, one could denote as a surface oxide, or ``stage I'' of oxide formation. This is to distinguish it from sub-surface oxygen, which often conveys more the notion of a randomly incorporated O species in the near-surface fringe. Although we expect from the deformation cost argument an enrichment of the latter compared to bulk dissolved O, one should stress that an experimental identification of either (bulk or sub-surface O) species will be very difficult: The concentration of bulk dissolved O is thermodynamically quite low in the late $4d$ TM crystal lattices \cite{reuter02a,todorova04} and will thus only yield small signatures (which as a note aside does not mean that the total amount of dissolved O in the whole crystal must be small though). And even though possibly enriched to a higher concentration, the signals from sub-surface O will also be small, since now only a small volume of the crystal will contribute. In fact, sub-surface oxygen has hitherto been quite elusive with only very few experimental studies claiming to have provided {\em direct} evidence for it. What is more frequent is that spectroscopic features that could not otherwise be explained have been {\em interpreted} as being due to a sub-surface oxygen species. 

Within the deformation cost picture, it appears in any case rather likely that (depending on the O exposure and kinetic barriers) a (lateral and vertical) aggregation of sub-surface oxygen will directly lead to ordered surface oxide structures. We will see below, that these structures are indeed oxide precursors, which may often have a somewhat similar stoichiometry as the bulk oxides (e.g. RuO$_2$ and Ag$_2$O for the two mentioned trilayer structures, respectively), but the geometries can differ significantly. Experimentally it may be similarly involved, though not impossible, to identify such structures, because also an ensuing transition to other oxidic precursors or eventually the ``true'' bulk oxide (film) may proceed fast.

\subsection{Formation of the bulk oxide}

\begin{figure}
\scalebox{0.38}{\includegraphics{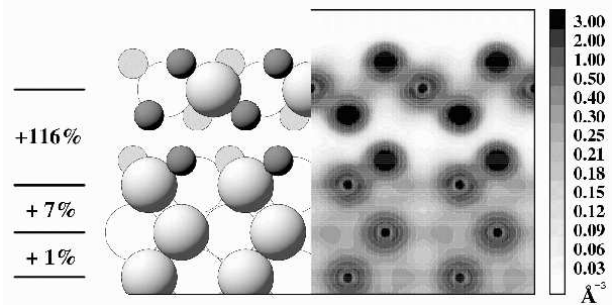}}
\caption{Side view of the detached ``floating trilayer'' of O at Ru(0001), corresponding to the most stable geometry with $\Theta = 3$\,ML. The left half of the figure shows the schematic geometry together with the relative layer expansions compared to the bulk distance and the right half the calculated electron density. Ru = large spheres, O = small spheres, Ru atoms not lying in the shown plane are whitened (from Ref. \onlinecite{reuter02b}).}
\label{float}
\end{figure}

Continued accommodation of oxygen at the surface will eventually yield to the phase transition towards the ``true'' bulk oxide structure. For different materials this happens at different oxygen load, and the pathway from the first formed surface oxide to the final bulk oxide may well comprise other oxidic precursor structures. In general, one has to admit that our atomic-scale knowledge specific to this step in oxide formation is again very shallow. Particularly for the more noble TMs the low stability of the bulk oxides requires either very high oxygen pressures and/or rather low temperatures for the experimental preparation. At the lower temperatures the growth is largely affected by kinetic limitations though, so that at Pd(111) at best small PdO clusters without apparent crystalline order were hitherto reported \cite{zheng00}, while at Ag(111) there exists literally no atomic-scale knowledge on the oxidation process beyond the formation of an ordered $p(4 \times 4)$ surface oxide (see below). This situation is far better for Ru(0001) and Rh(111), where it is at least known that the oxidation process ends with the formation of crystalline rutile-structured RuO$_2$(110) \cite{over00} and corundum-structured Rh$_2$O$_3$(0001) films \cite{gustafson04}, respectively.

This detailed knowledge of both initial state (the aforedescribed O-Ru-O trilayer) and final state (the rutile structured RuO$_2$(110) film) made it possible to elaborate for the first time on an atomistic pathway for the structural phase transition towards the bulk oxide \cite{reuter02b}: If the oxygen load at the Ru(0001) surface is increased beyond the 2\,ML coverage O-Ru-O trilayer, cf. Fig. \ref{subORu}, the surprising finding of the corresponding DFT-GGA study was that these additional oxygen atoms would also go preferentially between the first and second substrate layer, i.e. where already one full oxygen layer exists. The result is displayed in Fig. \ref{float}. It shows that a significant relaxation takes place such that the formed O-Ru-O trilayer is essentially floating on an O/Ru adsorbate system. With the trilayer decoupled, increasing the oxygen load further then gives rise to two O-Ru-O trilayers and so forth. Still, this is not yet a know Ru-oxide structure, neither with respect to the energy, nor the geometry. 

\begin{figure}
\scalebox{0.38}{\includegraphics{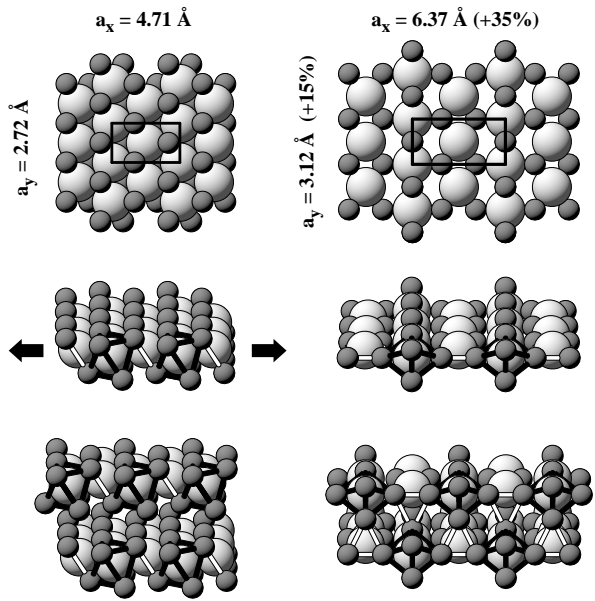}}
\caption{Atomic geometries of the O-Ru-O trilayer shown in Fig. \ref{float} (left-hand side) and of the final rutile RuO${}_2$(110) oxide structure (right-hand side). Top panel: top view, indicating the largely different sizes of the surface unit-cells. Middle panel: perspective view. The bulk-like rutile structure is achieved by simply stretching the trilayer in direction of the arrows, keeping the length of the drawn O-O bonds (white and black lines) rigid. This way an alternative sequence of fully coordinated Ru atoms (inside the tilted O octahedra, black bonds) and coordinatively unsaturated (cus) Ru atoms (surrounded by only four O atoms, white bonds) is created. Bottom panel: coupling of two trilayers vs. that of rutile layers. Whereas the self-contained trilayers hardly bind, new bonds can be formed in the RuO${}_2$(110) structure. The bridging O atoms of the full octahedra (black bonds) form the apex atoms of the new octahedra (white bonds) around the former cus Ru atoms. Note, that the `bonds' drawn do not refer to chemical bonds, but are merely used to guide the eye (from Ref. \onlinecite{reuter02b}).}
\label{accordion}
\end{figure}

As the trilayer on top has apparently only negligible influence on the underlying metal, a continued exposure to oxygen is likely to produce sequentially more and more of these trilayers, which one after the other will become decoupled from the metal substrate. This way a loosely coupled stack of trilayers will successively be formed. Comparing the geometry of these trilayers with the one of a rutile (110) layer, it can be noticed that the former can be transformed into the latter by a simple accordion-like lateral expansion of the trilayer as explained in Fig. \ref{accordion}. With this expansion the self-contained trilayer unfolds into a more open geometry, in which every second Ru atom is undercoordinated compared to the ideal sixfold oxygen surrounding both offered in the trilayer and in bulk RuO${}_2$. It is precisely these coordinatively unsaturated (cus) metal atoms that may eventually lead to a stabilization of the rutile structure for higher film thicknesses, as they can provide extra inter-layer bonds that can not be formed between the saturated trilayers. While for only one layer thick films the more compact trilayer offers therefore a more favorable configuration, this preference was already found reversed at a two layer thickness, eventually inducing the transition to rutile through the accordion-like expansion.\cite{reuter02b}

Admittedly, this idealized pathway for one specific surface can barely be able to represent the complex atomic rearrangements and stoichiometry changes one can expect more generally during this oxidation step. Still, it contains at least one aspect that appears quite crucial: the effect of strain. As already pointed out before, oxygen incorporation will always lead to a substantial local expansion of the crystal lattice of late $4d$ TMs, and known bulk oxide structures of the latter are accordingly much more open compared to the close-packed bulk metals. In the direction perpendicular to the surface, nothing opposes such an expansion and the corresponding vertical relaxations upon O incorporation can be quite substantial. Lateral relaxations, on the other hand, will conflict with the binding of the formed oxidic fringe to the underlying metal lattice. In the formation of (surface) oxides, the system therefore has to find an optimum energy compromise between elastic strain relaxation and good coupling to the underlying substrate. This may result in commensurable or incommensurable overlayers, and both types have already been observed experimentally on the late $4d$ basal surfaces.\cite{over00,gustafson04,lundgren02,carlisle00b} In addition, strain may be easier relaxed at more open sites like steps, pointing again at the relevance of the latter for the oxidation process. Apart from the strain energy, another influential factor is the thermodynamic driving force to form an oxide. The latter will roughly scale with the bulk oxide heat of formation, which steadily decreases from RuO$_2$ to Ag$_2$O. The concomitant, lower stability of oxidic structures {\em per se}, together with easier to deform metal lattices, might then render the coupling to the underlying substrate more important towards the more noble metals, i.e. the nature of the oxide-metal interface plays an increasing role. In this respect it is intriguing to notice that particularly at Pd(111) and Ag(111) pseudomorphic surface oxide structures have been reported, with a structure that seems particularly suited to maximize the interaction with the metal lattice below.

\section{Implications for oxidation catalysis}

\subsection{The role of the gas phase}

A salient feature emerging from studies of the kind briefly sketched in the first part of this review is that the initial oxidation of late $4d$ TMs proceeds via few atomic-layer thin oxidic structures, the electronic and geometric structure of which may deviate significantly from their known bulk oxide counterparts. This makes these surface oxide films distinct to surfaces of both bulk metals and bulk oxides, and one is led to wonder whether they exhibit specific new properties that can be exploited for a desired functionality like catalysis. In this respect, one has to recognize, however, that in such applications (and many experiments) the actual thermodynamic variable is not the oxygen accommodation at the surface, but the partial pressures and temperature in the surrounding gas phase. Instead of looking at the change of the state of the surface as e.g. a function of increased exposure (or uptake) as done in the preceding sections, the relevant question to ask is then rather what happens to the TM surface given certain environmental conditions? 

Exposed to an unlimited supply of gas phase particles characterized by the applied pressures and temperature, the surface will adapt on time scales set by the kinetic limitations. Already these time scales could be sufficiently long to render corresponding metastable states interesting for applications. In fact, the classic example is a slow thickening of oxide films due to limitations in the diffusion of oxygen atoms from the surface to the oxide-metal interface, or in the diffusion of metal atoms from the interface to the surface.\cite{fromhold76,roosendahl00} Directly at the surface a similar bottleneck can be the penetration of oxygen, which might be significantly facilitated at steps and defects and thus adds to their relevance particularly for the formation of surface oxides also from a kinetic point of view. Still, although such kinetic barriers might slow the oxidation process down beyond noticeable time scales, in a pure oxygen gas phase the true thermodynamic ground state will eventually be reached. With respect to oxide formation this is often loosely equated with a completely oxidized sample, and it thus came as quite a surprise that for the more noble $4d$ TMs DFT-GGA indicates that there are in fact $(T,p)$-conditions, for which a nanometer-thin surface oxide represents the thermodynamically stable phase. This means that in such environments a thin film is automatically formed at the oxidizing surface, the growth of which is self-limited to the potentially particularly interesting nanometer or sub-nanometer width.

Reactive multi-component gas phases, as typical for e.g. catalysis, add another twist to this situation. In this case, we are dealing with an open system, in which the supply of reactant gases comes into contact with the solid surface, where a chemical reaction produces a new substance that is then transported away. The average surface structure and composition is in this case entirely determined by the kinetics of the various underlying atomistic processes like adsorption, desorption, diffusion or reaction. Still, for given temperature and partial pressures of the reactants a steady-state situation may be reached (or even aspired for a stable catalyst performance), but this steady-state must not necessarily correspond to any thermodynamic state. Particularly for oxidation reactions, the O-rich environment might for example constantly oxidize the catalyst surface, while the on-going reaction continuously consumes oxygen in such a way, that oxide formation never proceeds beyond a certain state. Similar to the thermodynamic situation described above, this could also lead to a self-limited thickness of the oxidized surface fringe.

\subsection{Stability of surface oxides in an oxygen environment}

If we initially focus on the effect of a pure oxygen gas phase, it is useful to first resort to a thermodynamic description. This allows to assess the various stages in the oxidation process when the surface is fully equilibrated with its environment, and thus provides a sound basis for an ensuing discussion on how kinetic limitations could affect the picture. Particularly interesting in the context of this review is in this respect whether the thin surface oxides are mere kinetic precursors to bulk oxide growth or may actually represent a stable phase for certain gas phase conditions, as well as whether we can identify a trend over the late $4d$ TM series. For a surface in contact with a gas phase characterized by temperature $T$ and pressure $p_{O_2}$, a thermodynamic description takes place in the grand-canonical, constant $(T,p)$-ensemble and one needs to generalize eq. (\ref{bindeng}) to evaluate the Gibbs free energy of adsorption \cite{li03,michaelides03,reuter04b}
\begin{eqnarray}
\label{gibbs1}
\lefteqn{\Delta G(\mu_{\rm O}) \; =} && \\ \nonumber
&=& - \frac{1}{A} \left( G_{\rm O@surf.}
- G_{\rm surf.} - N_{\rm O} \mu_{\rm O} - \Delta N_{\rm M} \mu_{\rm M} \right) \quad.
\end{eqnarray}
Here, $A$ is the area of the surface unit-cell, and $G_{\rm O@surf.}$ and $G_{\rm surf.}$ are the Gibbs free energies of the oxidized and clean surface, respectively. $\mu_{\rm O}$ and $\mu_{\rm M}$ are the chemical potentials of oxygen and metal atoms, and $N_{\rm O}$ is the number of oxygen atoms contained in the oxidized surface structure for which $\Delta G(\mu_{\rm O})$ is computed. Finally, $\Delta N_{\rm M}$ is the difference in the number of metal atoms between the reference clean surface and the oxidized surface structural model.

\begin{figure*}[t]
\scalebox{0.4}{\includegraphics{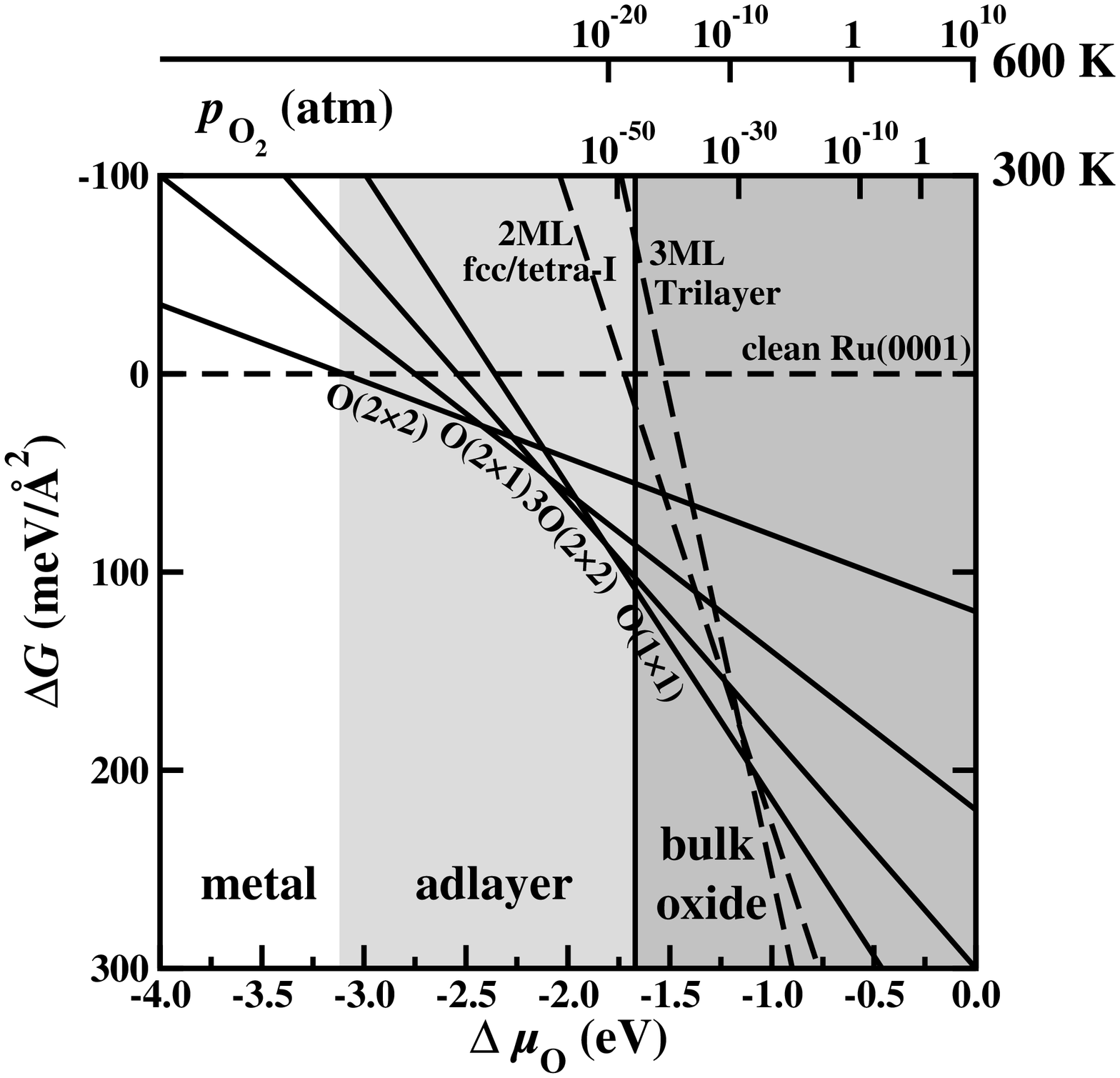}\quad\quad\quad\includegraphics{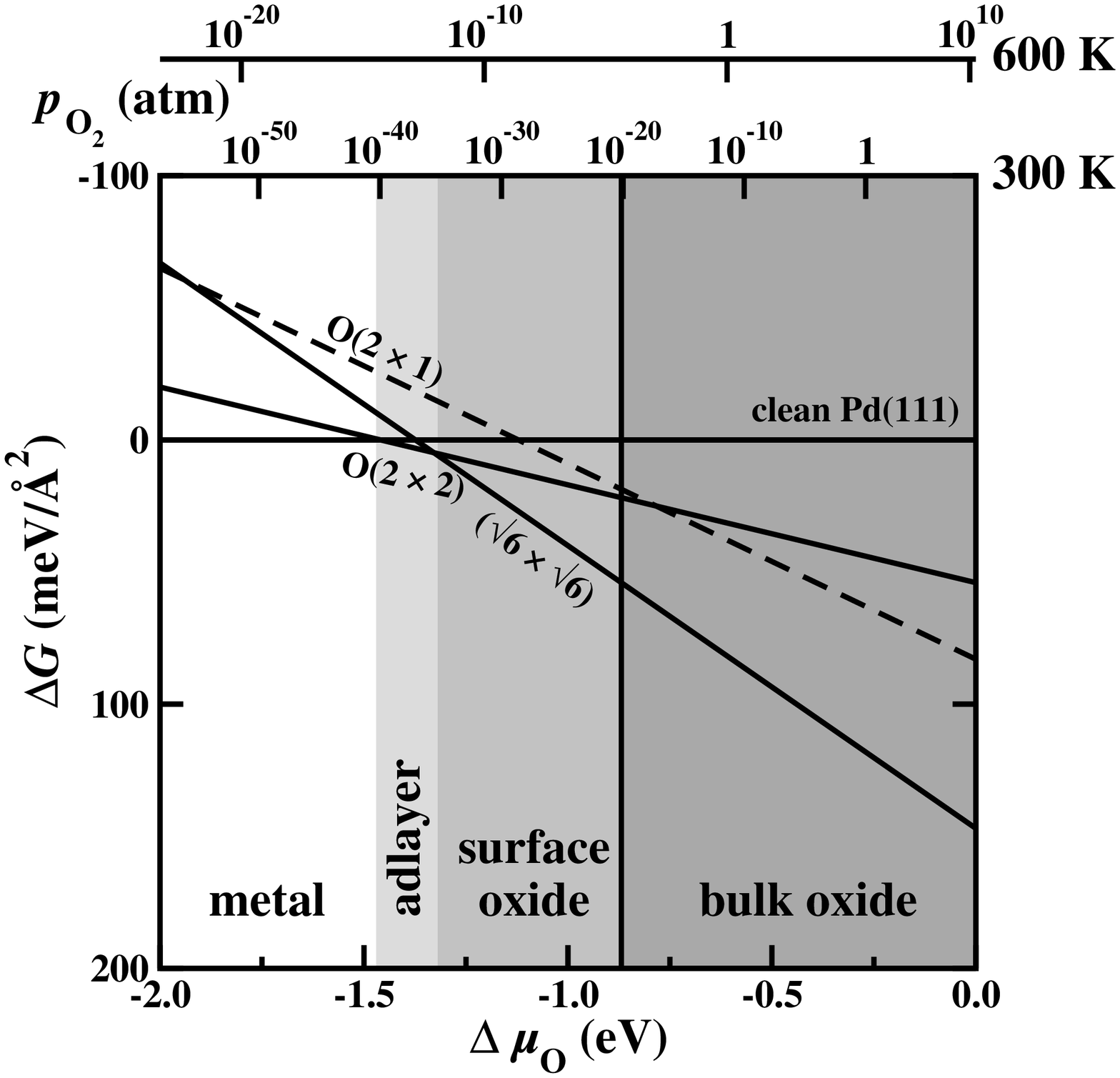}}
\caption{Computed DFT-GGA Gibbs free energy of adsorption for several ordered on-surface adsorbate structures, surface oxides and the bulk oxide. The dependence on $\Delta \mu_{\rm O}$ is also translated into pressure scales at $T=300$\,K and $T=600$\,K for clarity. In the bottom of the figure, the ``material type'' which is stable in the corresponding range of oxygen chemical potential is listed and indicated by the shaded regions. Left: Ru(0001). Compared are O$(2 \times 2)$, O$(2 \times 1)$, 3O$(2 \times 2)$ and O$(1 \times 1)$ adsorbate phases with O in hcp hollow sites, cf. Fig. \ref{miraprl1}, with the two O-Ru-O trilayer structures discussed in connection with Figs. \ref{subORu} and \ref{float}. Right: Pd(111). Compared are O$(2 \times 2)$ and O$(2 \times 1)$ adsorbate phases with O in fcc sites, cf. Fig. \ref{miraprl1}, with the so-called ``$(\sqrt{6} \times \sqrt{6})$'' surface oxide shown in Fig. \ref{pdo111} (the binding energies used to construct these graphs are taken from Refs. \onlinecite{reuter02b,lundgren02,reuter04b}).}
\label{deltaG}
\end{figure*}

In the difference between the two Gibbs free energies, the contributions due to vibrational free energy, configurational entropy and $pV$-term cancel to some extent, and for the oxidized late $4d$ TM surfaces discussed here replacing the Gibbs free energy difference by the difference of only the leading total energy terms appears to provide a reasonable approximation.\cite{xie99,reuter02c} This simplifies eq. (\ref{gibbs1}) to
\begin{eqnarray}
\label{gibbs2}
\lefteqn{\Delta G(\Delta \mu_{\rm O}) \; \approx} && \\ \nonumber
&\approx& \frac{1}{A} \left[ N_{\rm O} E_{\rm b}({\rm O@surf.}) + N_{\rm O} \Delta \mu_{\rm O} + \Delta N_{\rm M} \mu_{\rm M} \right] \quad,
\end{eqnarray}
where $E_{\rm b}({\rm O@surf.})$ is the average binding energy of oxygen in the oxidized surface, evaluated via eq. (\ref{bindeng}), and where the oxygen chemical potential is now measured with respect to the total energy of an isolated O$_2$ molecule as reference zero, $\Delta \mu_{\rm O} = \mu_{\rm O} - E_{\rm O_2}/2$. 

Eq. (\ref{gibbs2}) has a rather intuitive structure: Forming the oxidized surface by accomodating $N_{\rm O}$ oxygen atoms yields an energy gain of $N_{\rm O} E_{\rm b}({\rm O@surf.})$, that is opposed by the cost of taking these O atoms out of a reservoir, hence $N_{\rm O} \Delta \mu_{\rm O}$. The equivalent term $\Delta N_{\rm M} \mu_{\rm M}$ comes only into play for oxidized surfaces, where the total number of metal atoms is different to the one of the reference clean metal surface, and represents then the cost of transferring the corresponding number of metal atoms to or from a reservoir with chemical potential $\mu_M$. For the metal atoms, this reservoir is for oxidized surfaces often conveniently chosen to be the metal bulk, with which the surface is assumed to be in equilibrium.\cite{li03,michaelides03,reuter02c} The oxygen atom reservoir, on the other hand, is given by the surrounding gas phase, allowing to relate $\Delta \mu_{\rm O}$ to temperature $T$ and pressure $p_{\rm O_2}$ via ideal gas laws.\cite{reuter02c} For a given $\Delta \mu_{\rm O}(T,p_{\rm O_2})$ in the environment, the thermodynamically stable phase maximizes $\Delta G(\Delta \mu_{\rm O})$, and eq. (\ref{gibbs2}) can be used to compare structural models for possible oxidation states of the surface, e.g. oxygen adlayers with different coverage versus surface or bulk oxides. Since $E_{\rm b}({\rm O@surf.})$ from eq. (\ref{bindeng}) enters into eq. (\ref{gibbs2}), all the caveats about DFT calculations of the oxygen binding energy mentioned in the beginning hold here as well. This uncertainty is further aggravated by the additional approximation due to the neglected free energy contributions, and dictates utmost care in the interpretation of the obtained results. 

When explicitly comparing a set of structures in the sketched {\em atomistic thermodynamics} approach, surfaces with higher O content will become more favorable the more O-rich the environment. In the limit of low oxygen pressures the clean surface will obviously be most stable ($\Delta \mu_{\rm O} \rightarrow - \infty$, cf. eq. (\ref{gibbs2})), but this will eventually change with increasing chemical potential in the gas phase. The more oxygen is accommodated in a structure, the steeper the slope of the corresponding $\Delta G(\Delta \mu_{\rm O})$ curve with $\Delta \mu_{\rm O}$. In the limiting case of the bulk oxide with its infinite number of O atoms this will finally yield a vertical line at a value of the oxygen chemical potential that equals the bulk oxide heat of formation per O atom.\cite{reuter04b} For any higher value of $\Delta \mu_O$, the bulk oxide represents then the thermodynamically stable phase.

This overall structure is nicely visible in Fig. \ref{deltaG}, which compares several of the above discussed oxidation stages at the Ru(0001) surface. While for $\Delta \mu_{\rm O} < -3.1$\,eV the clean surface is most stable, four different O adsorbate phases with O in hcp sites and increasing coverages from 1/4\,ML up to 1\,ML become progressively more stable for more O-rich environments. Above $\Delta \mu_{\rm O} < -1.7$\,eV, bulk RuO$_2$ represents the most stable phase. The stability range of this bulk oxide is therewith so large that it in fact covers all realistically attainable pressures for temperatures below $T \sim 600$\,K, as illustrated in Fig. \ref{deltaG}. In particular, one also notices that the O-Ru-O trilayer structures discussed above as possible intermediates before the bulk oxide formation are never thermodynamic phases, suggesting that at best they could only be stabilized by kinetic limitations in the ensuing phase transition to the bulk oxide.

A qualitatively similar picture has recently been reported for Rh(111) \cite{gustafson04}, where after adsorbate phases with O in fcc sites, bulk Rh$_2$O$_3$ becomes most stable at higher oxygen chemical potentials. Due to the already decreased heat of formation per O atom of Rh$_2$O$_3$ compared to RuO$_2$, the stability range of the bulk oxide is smaller, but starts in contrast to the situation at Ru(0001) nevertheless before higher coverage adsorbate phases corresponding to $\Theta \stackrel{{\scriptscriptstyle >}}{{\scriptscriptstyle \sim}} 0.5$\,ML could become favorable. This limit is only approximate though, since one should keep in mind that in such atomistic thermodynamics plots only the stability of a given set of structures is compared, and there could well be yet uncharacterized, higher coverage structures with a sufficiently high average binding energy to show up as a new stable phase. Still, bulk oxide formation appears at this surface thermodynamically preferred already before a full ML O$(1 \times 1)$ adsorption phase, which is consistent with the trend towards easier O accommodation from Ru to Ag discussed in the first part of this review. Although a O-Rh-O trilayer structure equivalent to the one discussed for Ru(0001) above results similarly as only a kinetic precursor in DFT-GGA, it is interesting to note that at Rh(111) this structure could recently be experimentally prepared and characterized close to step edges.\cite{gustafson04}

\begin{figure}[t]
\scalebox{0.48}{\includegraphics{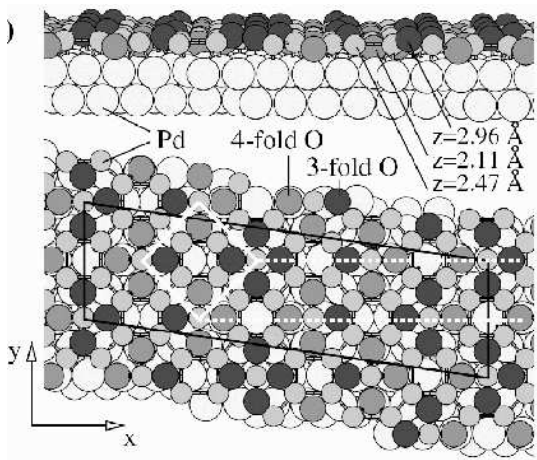}}
\caption{Top- and side view of the pseudo-commensurate, so-called ``$(\sqrt{6} \times \sqrt{6})$'' surface oxide structure determined on Pd(111). The atomic 
positions do not seem to resemble the bulk PdO arrangement, but exhibit roughly a O-Pd-O trilayer type structure (from Ref. \onlinecite{lundgren02}).}
\label{pdo111}
\end{figure}

Proceeding to Pd(111) one observes in most aspects a continuation of the identified trends. As apparent in Fig. \ref{deltaG}, the again reduced thermal stability of the PdO bulk oxide shows up by a further decreased stability range, which commences nevertheless already at an O chemical potential below the one required to stabilize higher on-surface adsorbate coverages with $\Theta \stackrel{{\scriptscriptstyle >}}{{\scriptscriptstyle \sim}} 1/4$\,ML. On the other hand, there is also a qualitatively new feature, since now the experimentally observed, so-called ``$(\sqrt{6} \times \sqrt{6})$'' 
pseudo-commensurate surface oxide structure displayed in Fig. \ref{pdo111} is not only a kinetic precursor, but represents the stable phase for an intermediate range of $\Delta \mu_{\rm O}$.\cite{lundgren02} This is therefore similar to the reported findings at Ag(111), of a $p(4 \times 4)$ surface oxide that is thermodynamically stable over a wide range of O chemical potentials, bounded on the one end by the stability range of a low coverage adsorbate phase and on the other end by the low-stability Ag$_2$O bulk oxide.\cite{li03,michaelides03} Both the ``$(\sqrt{6} \times \sqrt{6})$'' surface oxide on Pd(111) and the $p(4 \times 4)$ surface oxide on Ag(111) seem to have a highly complex, O-metal-O trilayered atomic structure that does not resemble the corresponding bulk oxides, cf. Fig. \ref{pdo111}.\cite{gustafson04,carlisle00a} Instead, these structures appear particularly suited to achieve a good coupling to the underlying substrate, and it is this extra contribution that enhances their stability over a wide $(T,p)$-range beyond the one of the known bulk oxides.

Summarizing the trend, we therefore see a changing role of thin surface oxide films in the oxidation process over the late $4d$ TM series. For Ru and Rh the present data suggests that such precursor structures to bulk oxide formation can only be stabilized by kinetic limitations. At the more noble metals Pd and Ag they instead represent a thermodynamically stable phase over an increasing range of gas phase conditions. In corresponding oxygen environments, these nanometer-thin films will eventually form on time scales set by possible kinetic limitations, but never grow thicker. Due to this finite thickness, the coupling at the oxide-metal interface and an atomic structure that can be quite different to the one of the known bulk oxides, one might suspect new properties that are distinct to those of surfaces of both bulk metals and bulk oxides, and could thus be of potential interest for applications. For the specific functionality in oxidation catalysis, one needs, however, also to assess the role played by the second reactant.

\subsection{``Constrained equilibrium''}

Although the average surface structure and composition is in catalysis even under steady-state conditions determined by the kinetics of the on-going elementary processes, it has been suggested that a thermodynamic approach with certain constraints could still provide a first insight into the effect of the reactive environment on the state of the catalyst surface.\cite{reuter03a,reuter03b} A straightforward generalization of the above described atomistic thermodynamics concept to a multi-component environment would be to account in eq. (\ref{gibbs2}) for a corresponding number of reservoirs described by chemical potentials $\Delta \mu_i$, where $i$ denotes the different gas phase species. Modeling the catalyst surface in the reactive environment, each reactant is thus represented by a corresponding reservoir, with which the surface is in contact. However, in a full thermodynamic description all reservoirs would also be mutually in equilibrium. Under catalytically interesting environmental conditions, this yields directly the products as most stable environment, and is thus inadequate to treat the effect of exposure to the reactant gas phase. 

Since it is precisely the high free energy barrier for the direct gas phase reaction that requires the use of the catalyst in the first place, it is appealing to instead consider the surface in contact with separate, independent reactant reservoirs. The surface would then be fully equilibrated with each reactant species, but the latter are not equilibrated with each other. Although neglecting the direct gas phase reaction is well justified, such a ``constrained equilibrium'' approach is still highly approximate, as it effectively also neglects the on-going catalytic product formation at the surface itself, i.e. in other words exactly what one is in principle out to study. In this respect one should stress that the whole concept of a ``constrained equilibrium'' is not designed to yield precise answers to the microscopic state and reactivity of the catalyst surface. For that, the very dynamic behavior must be modeled by statistical mechanics, as will be illustrated below. Yet, with atomistic thermodynamics the large scale behavior can be traced out first, possibly identifying those $(T,p_i)$-conditions where a more refined treatment explicitly accounting for kinetic effects is necessary.

\begin{figure}
\scalebox{0.32}{\includegraphics{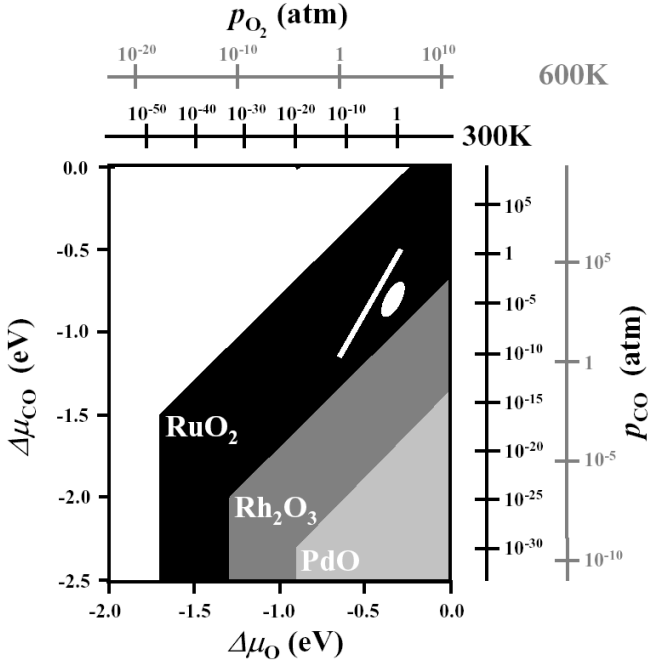}}
\caption{Stability regions of bulk oxides of the late $4d$ TMs in
$(\Delta \mu_{\rm O}, \Delta \mu_{\rm CO})$-space, as computed with DFT-GGA. Additionally, pressure scales are drawn for $T = 300$\,K and $T = 600$\,K.
The white line indicates environmental conditions corresponding 
to $p_{\rm O_2} = p_{\rm CO} = 1$\,atm and 300\,K $\le T \le$\,600\,K, 
and environments relevant for CO oxidation catalysis over these 
metals would correspond to the near vicinity of this line. More 
specifically, the small white area indicates the $(T,p)$-conditions 
employed in a recent experimental study by Hendriksen, Bobaru and 
Frenken on Pd(100) (s. text) \cite{hendriksen04} (from Ref. \onlinecite{reuter04b}).}
\label{bulkoxidestab}
\end{figure}

How this works in practice is nicely illustrated by comparing the stability range of the bulk oxides of the late $4d$ TM series within the context of CO oxidation catalysis, i.e. in a gas phase formed of oxygen and CO.\cite{reuter04b} The progressively decreasing heats of formation of the more noble metal oxides already discussed in the last section are also clearly discernible in Fig. \ref{bulkoxidestab}, which shows the oxide stability range as a function of the two chemical potentials in the gas phase. At very low CO chemical potential, the situation in a pure O$_2$ surrounding is recovered, and the stability range of RuO$_2$ and PdO coincides with the corresponding bulk oxide shaded regions in Fig. \ref{deltaG}. With increasing CO content in the gas phase, this picture changes and the stability range gets eventually reduced due to CO-induced decomposition. Although uncertainties in the modeling of this decomposition translate presently into rather large error bars in the respective diagonal limits in Fig. \ref{bulkoxidestab},\cite{reuter04b} a clear trend still emerges: Only the stability range of bulk RuO${}_2$ extends well into environmental conditions representative of technological CO oxidation catalysis, roughly indicated in Fig. \ref{bulkoxidestab} as a white line. For Rh on the other hand the stability range of its oxide has already decreased significantly. Even if the real system was in a situation close to thermodynamic equilibrium, either metal or bulk oxide could prevail under ambient conditions, depending on whether the exact partial pressures and temperatures in the gas phase correspond to a situation above or below the stability line (which is at present impossible to assess due to the large uncertainty in the calculated position of this line). The bulk oxides of Pd and of Ag are finally already so unstable that they are unlikely to play a role in CO oxidation catalysis. The stability of Ag${}_2$O in fact being so low, that it does not even show up anymore within the chemical potential range plotted in Fig. \ref{bulkoxidestab}.

\begin{figure*}
\scalebox{0.9}{\includegraphics{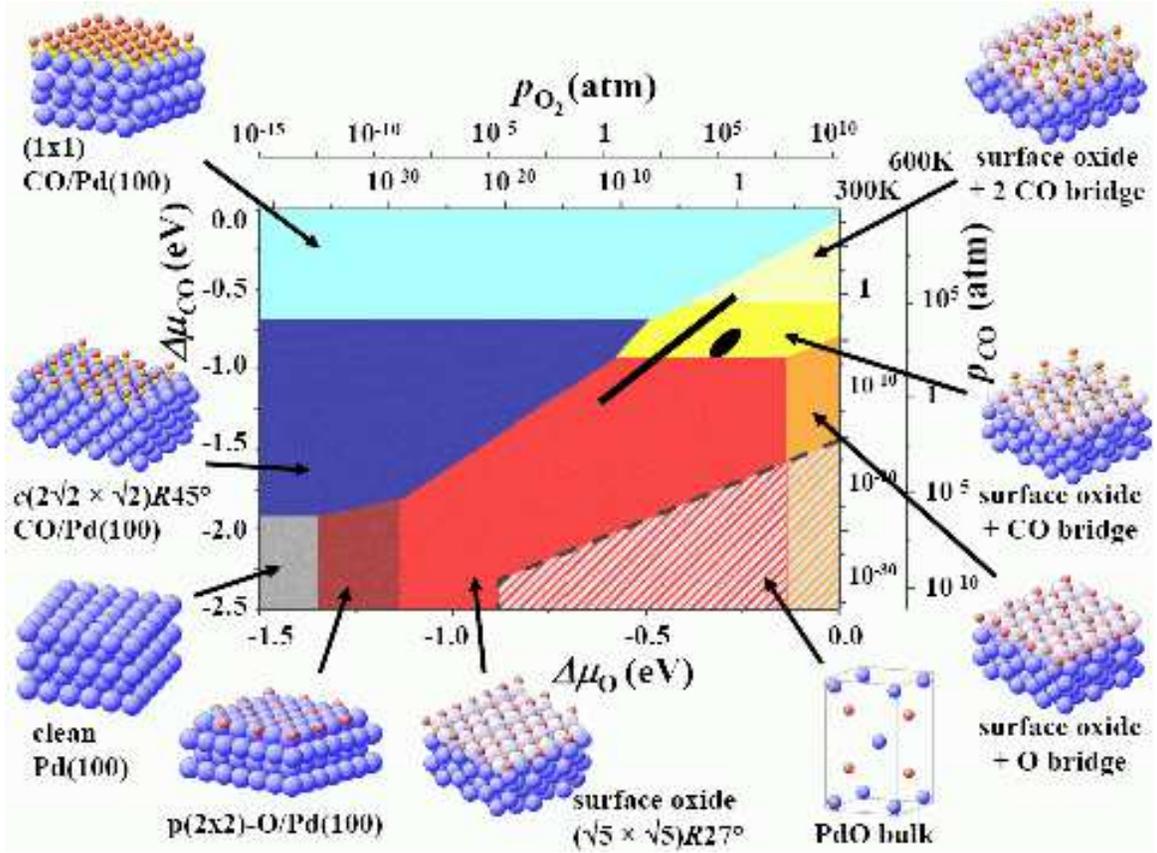}}
\caption{DFT-GGA surface phase diagram of stable surface structures at a Pd(100) surface in ``constrained equilibrium'' with an O$_2$ and CO environment. The stability range of bulk PdO is the same as the one shown in Fig. \ref{bulkoxidestab}. Phases involving the $(\sqrt{5} \times \sqrt{5})R27^{\circ}$ surface oxide exhibit a largely increased stability range, that now comprises gas phase conditions representative of technological CO oxidation catalysis ($p_{\rm O_2} = p_{\rm CO} = 1$\,atm and 300\,K $\le T \le$\,600\,K, marked by the black line as in Fig. \ref{bulkoxidestab}) (from Ref. \onlinecite{rogal04}).}
\label{pdosurf1}
\end{figure*}

It is, however, important to realize that already quite thin oxide films at the surface would be sufficient to induce a changed catalytic functionality. In this respect, just considering the stability of bulk oxides as representatives of thick, bulk-like oxide overlayers is not enough. Particularly at the more noble metals, we discussed already the extended stability range of the few 
atomic-layer thin surface oxide structures in a pure O$_2$ gas phase, which could similarly prevail in a reactive multi-component environment. While the results of Fig. \ref{bulkoxidestab} indicate that thick, bulk-like PdO and Ag$_2$O overlayers will hardly play a role in CO oxidation catalysis, the picture could be quite different for the nanometer thin surface oxide structures. Fig. \ref{pdosurf1} exemplifies this with corresponding results for the case of a Pd(100) surface in constrained equilibrium with an O$_2$ and CO environment.\cite{rogal04} Similar to the aforediscussed situation at Pd(111), oxidation proceeds also at Pd(100) via a trilayer-structured O-Pd-O surface oxide film \cite{zheng00}, that was recently identified as essentially a strained and rumpled layer of PdO(101) on top of the Pd(100) substrate \cite{todorova03}. Although the (101) orientation is not a low-energy surface of bulk PdO,\cite{rogal04b} it gets stabilized in the commensurate $(\sqrt{5} \times \sqrt{5})R27^o$ arrangement of the thin oxide film due to a strong coupling to the underlying metal substrate. In line with our previous discussion one would then expect an extended stability range compared to bulk PdO, that was indeed subsequently found both theoretically and experimentally for a pure O$_2$ gas phase.\cite{lundgren04} The atomistic thermodynamics results summarized in Fig. \ref{pdosurf1} suggest furthermore that this would indeed hold equally for Pd(100) in an O$_2$ and CO surrounding. Comparing the stability of a large set of on-surface (co)adsorption, surface oxide and bulk oxide structures in constrained equilibrium with the reactive environment \cite{rogal04}, several phases involving the $(\sqrt{5} \times \sqrt{5})R27^{\circ}$ surface oxide are found to be most stable over a wide range of $(T,p_{\rm O_2},p_{\rm CO})$-conditions that largely exceed the small stability range of bulk PdO. 

The range extends in fact so much that it even just comprises the gas phase conditions typical for technological CO oxidation, i.e. partial pressures of the order of 1\,atm and temperatures around 300-600\,K. Similar conditions were e.g. also chosen in a recent {\em in-situ} reactor scanning tunneling microscopy (STM) experiment of CO oxidation at Pd(100) by Hendriksen, Bobaru and Frenken\cite{hendriksen04}, cf. Figs. \ref{bulkoxidestab} and \ref{pdosurf1}. With increasing O$_2$:CO partial pressure ratio, the authors observed substantial changes in the surface morphology, which they assigned to the formation of a thin oxidic overlayer \cite{hendriksen04}. Inspecting Fig. \ref{pdosurf1} it is tempting to identify this overlayer as the already characterized $(\sqrt{5} \times \sqrt{5})R27^{\circ}$ surface oxide, but this would be rushing beyond what can be safely concluded on the basis of the presently available DFT data: First, surface structures that are not yet explicitly compared in the atomistic thermodynamics approach could still change the surface phase diagram. Second, kinetic effects could significantly affect the oxidation state of the surface. This is the case, when the on-going catalytic reaction consumes surface species faster than they can be replenished from the surrounding gas phase, thus yielding an average surface structure and composition that can be quite distinct to the predictions obtained within the constrained equilibrium concept. It could in particular be that a fast consumption of surface oxygen atoms reduces the stability range of the surface oxide well below the technologically interesting environmental conditions (which are even in constrained equilibrium quite close to the transition to the CO-covered Pd(100) surface phases). At the moment, the only safe take-home message from the data presented in Fig. \ref{pdosurf1} is therefore that instead of the already ruled-out thick bulk-like PdO film, the nanometer thin $(\sqrt{5} \times \sqrt{5})R27^{\circ}$ surface oxide appears as a likely candidate to actuate the increased catalytic activity connected with the morphology changes reported by Hendriksen and coworkers.\cite{hendriksen04}

Still, the proximity of the technologically relevant high-pressure gas phase conditions to the phase transition between surface oxide and CO-covered Pd(100) in Fig. \ref{pdosurf1} is intriguing and stimulates further speculations. Depending on the exact partial pressures and temperatures, either the surface oxide or the Pd(100) metal surface could prevail. Even oscillations between the two phases are conceivable as a consequence of fluctuations, when operating under environmental conditions very close to the stability limit. At finite temperatures, configurational entropy may furthermore give rise to a complex phase coexistence at the surface,\cite{reuter02c,reuter03b} which would then go in the direction of what was in the {\em in-situ} reactor STM experiment interpreted as a continuous consumption and reformation of the surface oxide even under the employed stationary reaction conditions.\cite{hendriksen04} Instead of the one-time formation of a catalytically active oxide film in the case of Ru, this would point at a much more dynamic state of the Pd surface in the reactive CO oxidation environment: The catalytic activity could be intimately connected to a reversible (local) formation and reduction of the thin surface oxide, possibly even giving rise to spatio-temporal pattern formation phenomena as often reported for CO oxidation at Pt-group metals.\cite{hendriksen04,bondzie96,hendriksen03a} In such cases, also the domain boundaries between surface oxide and adlayers could be of particular importance for the overall reactivity.

\subsection{Kinetically limited film thickness}

The constrained equilibrium description discussed up to now conveys the impression that a possible oxidation of the catalyst surface in the O-rich environments of oxidation catalysis would rather yield bulk-like thick oxide films on Ru, but thin surface oxide structures on the more noble $4d$ metals. This reflects the decreasing heat of formation of the bulk oxides over the late TM series, and seems to suggest that it is primarily at Pd and Ag where oxide formation in the reactive environment could be self-limited to nanometer or sub-nanometer thin overlayers. Particularly for the case of Ru, Fig. \ref{bulkoxidestab} shows that the gas phase conditions typical for technological CO oxidation catalysis fall deep inside the stability regime of the bulk oxide, indicating that thermodynamically nothing should prevent a continued growth of the once formed oxide film. 

\begin{figure}
\scalebox{0.49}{\includegraphics{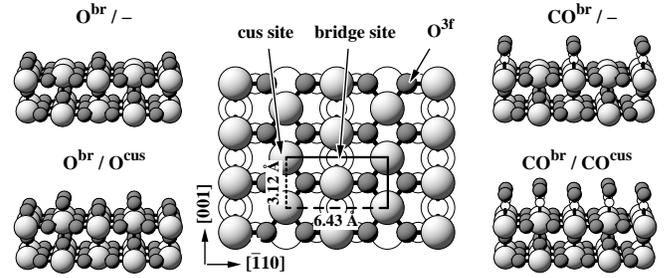}}
\caption{\label{ruo2_sites}
Top view of the RuO$_2$(110) oxide surface explaining the location of the two prominent adsorption sites (coordinatively unsaturated, cus, and bridge, br). Also shown are perspective views of the four steady-state adsorption phases present in Fig. \ref{ORukMC} (Ru = light large spheres, O = dark medium spheres, C = white small spheres).}
\end{figure}

The formation of crystalline, bulk-like RuO$_2$(110) during high-pressure CO oxidation catalysis has indeed been observed experimentally at Ru(0001), but it is interesting to note that even after long operation times the film thicknesses never exceeded about 20\,{\AA} \cite{boettcher97,over00,over04}. Generally, one could interpret this finding as reflecting kinetic limitations to a continued growth, presumably due to slow diffusion of either O or Ru atoms through the formed film.\cite{fromhold76,roosendahl00} While this could slow the thickening of the oxide overlayer down beyond hitherto measured time scales, it could also be that the kinetics of the on-going catalytic reaction affect the surface populations in such a way, that no further net diffusion through the film occurs and the nanometer thin oxide overlayer represents in fact a kinetically limited
steady-state in the given reactive environment. This would e.g. be the case, if the oxidation reactions at the surface consume adsorbed oxygen atoms at such a pace that on average O penetration into the film occurs not more frequently than the filling of surface sites with oxygen from inside the film. 

\begin{figure}
\scalebox{0.46}{\includegraphics{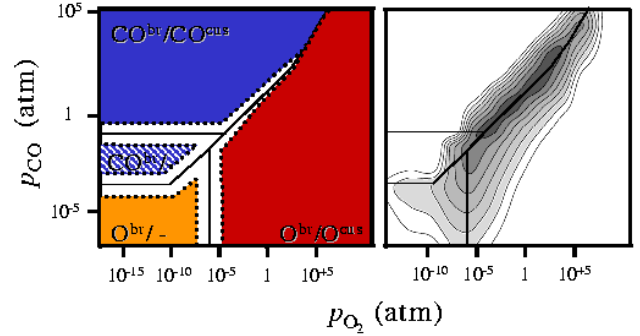}}
\caption{\label{ORukMC}
Left panel: Steady state surface structures of RuO$_2$(110) in an O$_2$/CO environment obtained by first-principles kMC calculations at $T = 600$\,K. In all non-white areas, the average site occupation is dominated ($> 90$\,\%) by one species, and the site nomenclature indicates the species at the two prominent adsorption sites offered by the surface: br = bridge sites, cus = coordinatively unsaturated sites, cf. Fig. \ref{ruo2_sites}. Right panel: Map of the corresponding catalytic CO oxidation activity measured as so-called turn-over frequencies (TOFs), i.e. CO$_2$ conversion per ${\rm cm}^{2}$ and second: White areas have a TOF\,$< 10^{11} {\rm cm}^{-2}{\rm s}^{-1}$, and each increasing gray level represents one order of magnitude higher activity. The highest catalytic activity (black region, TOF $> 10^{17}\,{\rm cm}^{-2}{\rm s}^{-1}$) is narrowly concentrated around the phase coexistence region, in which O and CO compete for both br and cus sites (from Ref. \onlinecite{reuter04c}).}
\end{figure}

To explore this in more detail one needs to go beyond the constrained equilibrium description and explicitly take kinetic effects into account in the theoretical modeling. This can for example be achieved via kinetic Monte Carlo (kMC) simulations, which can essentially be viewed as coarse-grained molecular dynamics simulations, following the time evolution of the system on the basis of the rates of all individual elementary processes occurring in the system.\cite{reuter04a,landau02,kang95} Evaluating the statistical interplay between this manifold of atomic-scale processes, kMC simulations can easily cover sufficiently long time scales to arrive at proper averages of quantities like surface populations, even if the system is not in equilibrium, but only in a steady-state as typical for catalysis. Such a modeling was recently performed for the CO oxidation over RuO$_2$(110), using DFT-GGA based rates for 26 elementary surface processes including adsorption, desorption, diffusion and reaction events.\cite{reuter04c} Of particular interest for the film thickness issue is the occupation of surface sites with O atoms in catalytically relevant gas environments. In this respect, the RuO$_2$(110) surface offers two prominent adsorption sites, commonly denoted as bridge (br) and coordinatively unsaturated (cus) sites as explained in Fig. \ref{ruo2_sites}.\cite{reuter02c} 

\begin{figure}
\scalebox{0.6}{\includegraphics{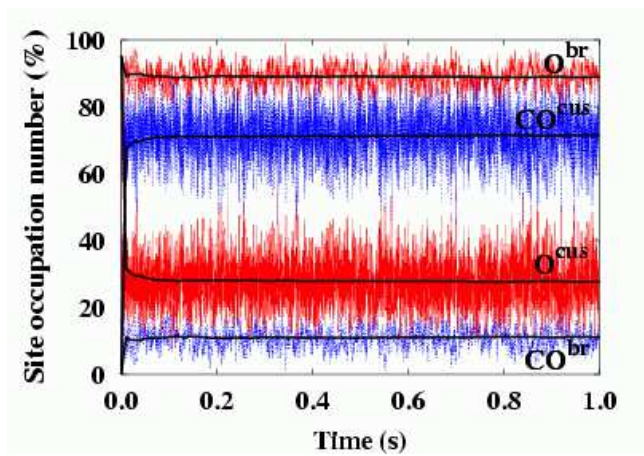}}
\caption{\label{ORuoccupation}
Time evolution of the site occupation by O and CO of the two prominent adsorption sites at the RuO$_2$(110) model catalyst surface shown in Fig. \ref{ruo2_sites}. The temperature and pressure conditions chosen ($T=600$\,K, $p_{\rm CO} = 20$\,atm, $p_{\rm O_2} = 1$\,atm) correspond to optimum catalytic performance, cf. Fig. \ref{ORukMC}. Under these conditions kinetics builds up a steady-state surface population in which O and CO compete for either site type at the surface, as reflected by the strong fluctuations in the site occupations (from Ref. \onlinecite{reuter04c}).}
\end{figure}

Figure \ref{ORukMC} shows the obtained steady-state average surface populations at $T=600$\,K as a function of the gas-phase partial pressures, together with the simultaneously computed turnover frequencies (TOF, in units of formed CO$_2$ per cm$^{2}$ per second). The latter result from evaluating the average occurrence of the reaction events over long time periods as a measure of the catalytic activity, and are intriguingly peaked around a narrow range of gas phase conditions. Contrary to the situation at most other partial pressures, the site occupation at br and cus sites is then not dominated by one species, i.e. either O or CO (or vacant). Instead, the kinetics builds up a surface population in which O and CO compete for either site type at the surface. This competition is in fact nicely reflected by the huge fluctuations in the surface populations with time apparent in Fig. \ref{ORuoccupation}. The dynamics at the surface in this ``active state'' is furthermore extremely fast, and the average time adsorbed O atoms stay in br or cus sites before desorbing or being reacted away is only of the order of fraction of milliseconds. Remarkably, it is particularly the concentration of O at the surface that turns out significantly lower than predicted within the constrained equilibrium approach \cite{reuter03a,reuter03b}. Even at the bridge sites, where O atoms bind very strongly \cite{reuter02c}, the O occupation is only 90\% and not 100\% as in the normally stable surface termination. This is quite surprising, and indicates already how effectively the on-going reaction consumes surface O species. It could therefore well be that the surface kinetics indeed limits a continued growth of the formed oxide film under this optimum catalytic performance.

\subsection{Surface oxidation and Sabatier principle}

The partial pressures and temperatures leading to this high activity state of the RuO$_2$(110) surface in the calculations, and even the absolute TOF values under these conditions, agree extremely well with detailed experimental studies measuring the steady-state activity in the temperature range from 300-600\,K and both at high-pressures and in UHV \cite{cant78,peden86,fan01}. This statement comprises already the emerging consensus that this Ru oxide also forms at the surface of Ru model catalysts in high-pressure environments \cite{reuter04c,over04}, i.e. the high activity that was originally ascribed to the Ru substrate was in fact due to RuO$_2$ \cite{cant78,peden86}. With this understanding a puzzle is finally resolved that had bothered surface scientists for quite a while: For a long time the high CO oxidation reactivity of supported Ru catalysts \cite{cant78}, as well as of Ru(0001) single crystals \cite{peden86} at high-pressures stood in sharp contrast to exceedingly low turnover rates observed under UHV conditions \cite{madey75}, thus representing a nice example of a so-called ``pressure gap'' system \cite{stampfl02}. 

Looking at the computed binding energies of oxygen at Ru(0001) shown in Fig. \ref{miraprl1} the low catalytic activity of this surface appears in fact quite plausible. According to the early Sabatier principle \cite{ertl97} high activity would in general rather be expected for intermediate bond strengths at the surface (stable enough to form, but weak enough to yield the final product). In this respect the computed medium binding energies at e.g. Rh(111) and Pd(111) in Fig. \ref{miraprl1} conform much better with the widespread use of these metals in catalytic car converters, whereas the oxygen bonding at Ru(0001) seems simply too strong. The correspondingly somehow enigmatic high catalytic activity of Ru(0001) at ambient pressures only became comprehensible when experimental studies reported the formation of bulk-like RuO${}_2$(110) films at the surface in such reactive environments \cite{boettcher97,over00}. The subsequently established \cite{kim01b} and above described high activity of this oxide surface then not only explained the pressure gap difference (Ru metal in UHV = low activity, RuO${}_2$ at high-pressures = high activity), but also reconciled this system with the Sabatier principle: The RuO${}_2$(110) surface indeed features oxygen species that are bound with intermediate bond strength comparable to the one of O at Rh(111) and Pd(111).\cite{reuter02c,fan01}

An equivalent, though reverse explanation has also been suggested for the most noble of the late $4d$ metals.\cite{li03b,li03c} According to Fig. \ref{miraprl1} oxygen binds to Ag(111) only very weakly, which in view of the Sabatier principle is again difficult to combine with the known importance of Ag as catalyst for several oxidation reactions. However, the bond strength of oxygen in the p$(4 \times 4)$ surface oxide is substantially higher and would therefore well fall into the more favorable medium binding energy range.\cite{li03c} If the surface oxide instead of Ag metal was the actually active surface in the high-pressure oxidation catalysis, the Sabatier principle would finally be reconciled with the complete trend of oxygen binding over the late $4d$ series apparent in Fig. \ref{miraprl1}: Both the too strong O binding on Ru and the too weak O binding on Ag is alleviated by the formation of oxides, which exhibit a similarly intermediate bond strength as oxygen at Rh or Pd. 

It thus almost appears as if the interaction of oxygen with TM catalysts would serve to ``tune'' the O-metal bond strength for optimum reactivity through an appropriate amount of surface oxidation. Since the thermal stability of bulk Ag$_2$O is already too low and in line with our preceding extended stability discussion, this oxidation can at Ag at best correspond to the formation of a thin surface oxide though. Yet, compared to the surface oxides at Pd, even the latter is probably not stable enough to play an active role in CO oxidation. This can be different in the less reducing environments of partial oxidation reactions, consistent with the predominant use of Ag catalysts for ethylene epoxidation or formaldehyde synthesis \cite{ertl97}.

These examples underline therefore already nicely the potential relevance of surface oxidation in the reactive surroundings of oxidation catalysis. In all cases discussed so far, this had a beneficial effect on the overall catalytic performance, but detrimental effects due to the formation of an inactive oxide film are of course equally well conceivable. Such a catalyst deactivation through oxidation has e.g. been frequently suspected in the experimental literature concerning Rh surfaces.\cite{oh83,kellog85,peden88} In other cases, the catalytic activity of oxide and metal surface might be comparable, so that surface oxidation would likely remain unnoticed unless explicitly looked for by experimental techniques sensitive to the atomic-scale.

\section{Conclusion}

\begin{figure}
\scalebox{0.42}{\includegraphics{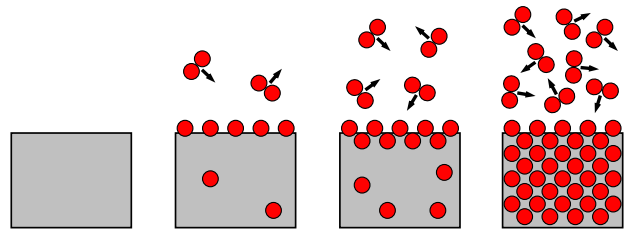}}
\caption{\label{oxischematic}
Cartoon sideviews illustrating the effect of an increasingly oxygen-rich atmosphere on a late TM metal surface. Whereas the clean surface prevails in perfect vacuum (left), finite O${}_2$ pressures in the environment first lead to oxygen adsorption phases. Apart from some bulk dissolved oxygen, the lower deformation cost will at increasing pressures lead to a preferential accommodation of oxygen in the near-surface fringe. Thin surface oxide structures are the salient consequence, before eventually thickening of the oxide film and formation of an ordered bulk compound set in. A thermodynamic or kinetic stabilization of the nanometer thin surface oxide could lead to novel functionalities, different from both bulk metal or bulk oxide surfaces.}
\end{figure}

Novel experimental and theoretical approaches addressing the effect of realistic environments on surface structure and composition have significantly increased the atomic-scale understanding of the oxide formation process of late transition metals over the last years. Some identified key features have been reviewed for the late $4d$ sequence from Ru to Ag, and are summarized schematically again in Fig. \ref{oxischematic}: A progressively more O-rich gas phase leads first to on-surface adsorption, followed by O penetration into the surface at increasing O accommodation. The incorporation takes place after the formation of dense oxygen overlayers in the case of Ru, but already at lower on-surface coverages towards the more noble metals. This results from the opposing trends in decreasing O-metal bond strength and decreasing deformation cost of the metal lattice, the latter being a consequence of the substantial local expansion necessarily connected to O incorporation into the close-packed TM lattices. The deformation cost is also responsible for an enrichment of oxygen in the easier to relax immediate surface fringe, eventually inducing the formation of thin oxidic structures at the surface. Although some bulk dissolution will always occur on entropy grounds, there is thus primarily a propensity to form few atomic-layer thin surface oxides, which mostly goes hand in hand with substantial structural changes. The resulting geometry is then not only characterized by an optimum binding within the more open compound structure, but can also be largely affected by the coupling to the underlying metal substrate (i.e. the nature of the oxide-metal interface) and residual strain. As consequence, complex atomic arrangements not resembling surfaces of bulk oxides or bulk metals have been reported for this step in the oxidation process. Only after continued oxygen accommodation, the thickening oxide film will eventually switch to a bulk oxide structure, and the continued film growth will then be more and more limited by the slow diffusion of O and/or metal atoms between surface and interface.

A most intriguing aspect of this surface oxidation is that it proceeds via the few atomic-layer thin surface oxide structures. This thickness can be ruled by kinetics, but at Pd and Ag it even seems that there are thermodynamic conditions for which the surface oxides represent the most stable phase. In corresponding environments such films are formed automatically, but the growth is self-limited to the potentially particularly interesting nanometer or sub-nanometer width. If stabilized in applications, the often quite distinct geometries of these films already suggest that new properties and functionalities might emerge that are not scalable from the known bulk materials. At present, very little to nothing is known about these properties, and one should recall that not even the corresponding bulk oxides of the late $4d$ TMs have hitherto received a lot of attention in atomic-scale studies. This opens up a wide new field for investigations of this new ``material class'' and its functionalities in various applications. 

With respect to oxidation catalysis, oxide films may even form as a natural consequence of the reactive environment, and since maybe hitherto unnoticed, they might then actuate some of the catalytic behavior that has traditionally been ascribed to the metal substrates. In this respect, the decreasing heat of formation leads to a clear trend for the bulk oxides: A bulk-like RuO$_2$ film can even sustain the reducing environments of high-pressure total oxidation reactions, whereas the feeble Ag$_2$O will hardly prevail. This is, however, not enough to exclude the relevance of oxidized structures for oxidation catalysis over Pd and Ag catalysts. Also due to the coupling to the underlying substrate, the thin surface oxides exhibit a much extended stability range, and it seems at the moment at least feasible that for example CO oxidation over Pd is actuated by them, or their constant consumption and reformation. The present data suggest a similar role for surface-confined oxidic structures at Ag, but only in less reducing environments e.g. typical for partial oxidation. This would then also offer an intuitive explanation for the often reported high catalytic activity of noble metals for such reactions. Together with the formation of highly active, nanometer thin RuO$_2$ films at the surface of Ru model catalysts, this underlines the potential importance of surface oxidation in the reactive environments of oxidation catalysis over late TMs. Although often not yet fully conclusive, already the reviewed new insights into the oxidation behavior help sometimes to reconcile the data with more general concepts like the Sabatier principle. Much more often, however, they lead to the falldown of ``established'' rules and preconceptions, which is a prospect that makes working in this field exciting and all the worthwhile.

\quad\\

\section*{Acknowledgments}

I gratefully acknowledge the valuable contributions and help from my (former) colleagues working on metal oxidation in the Theory Department of the Fritz-Haber-Institut: J\"org Behler, Veronica Ganduglia-Pirovano, Wei-Xue Li, Angelos Michaelides, Jutta Rogal, Cathy Stampfl and Mira Todorova. Particular thanks go to Matthias Scheffler for his continued support and for the insightful discussions on the intriguing facets of surface oxidation and catalysis.


\begin{references}

\bibitem{haruta97}
M. Haruta, Catalysis Today {\bf 36}, 153 (1997).

\bibitem{dreizler90}
R.M. Dreizler and E.K.U. Gross, {\em Density-Functional Theory},
Springer, Berlin (1990).

\bibitem{parr94}
R.G. Parr and W. Yang, {\em Density-Functional Theory of Atoms and Molecules},
Oxford University Press, Oxford (1994).

\bibitem{scheffler00}
M. Scheffler and C. Stampfl, {\em Theory of Adsorption on Metal Substrates}, In:
Handbook of Surface Science, Vol. {\bf 2}: Electronic Structure, K. Horn and M. 
Scheffler (Eds.), Elsevier, Amsterdam (2000).

\bibitem{gross02}
A. Gro{\ss}, {\em Theoretical Surface Science - A Microscopic Perspective},
Springer, Berlin (2002).

\bibitem{reuter04a}
K. Reuter, C. Stampfl and M. Scheffler, {\em Ab initio Atomistic Thermodynamics and Statistical Mechanics of Surface Properties and Functions}, In: Handbook of Materials Modeling, Vol. {\bf 1}: Fundamental Models and Methods, S. Yip (Ed.), Kluwer (in press).

\bibitem{behler-2004} 
J. Behler, S. Lorenz, K. Reuter, M. Scheffler, and B. Delley, ({\em to be published}).

\bibitem{perdew-PRL-1996}
J.P. Perdew, K. Burke and M. Ernzerhof, Phys. Rev. Lett. {\bf 77}, 3865 (1996).

\bibitem{zhang-PRL-1998}
Y. Zhang and W. Yang, Phys. Rev. Lett. {\bf 80}, 890 (1998).

\bibitem{perdew-PRL-1999} 
J.P. Perdew, S. Kurth, A. Zupan, and P. Blaha,
Phys. Rev. Lett. {\bf 82}, 2544 (1999).

\bibitem{filippi-PRL-2002}
C. Filippi, S.B. Healy, P. Kratzer, E. Pehlke, and M. Scheffler,
Phys. Rev. Lett. {\bf 89}, 166102 (2002).

\bibitem{surnev85}
L. Surnev, G. Rangelov and G. Bliznakov,
Surf. Sci. {\bf 159}, 299 (1985).

\bibitem{pfnur89} 
H. Pfn\"ur, D. Held, M. Lindroos, and D. Menzel,
Surf. Sci. {\bf 220}, 43 (1989).

\bibitem{comelli98}
G. Comelli, V.R. Dhanak, M. Kiskinova, K.C. Prince, and R. Rosei, 
Surf. Sci. Rep. {\bf 32}, 165 (1998).

\bibitem{stampfl-PRB-1996}
C. Stampfl and M. Scheffler, 
Phys. Rev. B {\bf 54}, 2868 (1996).

\bibitem{stampfl-PRL-1999}
C. Stampfl, H.J. Kreuzer, S.H. Payne, H. Pfn\"ur, and M. Scheffler, 
Phys. Rev. Lett. {\bf 83}, 2993 (1999).

\bibitem{ganduglia99}
M.V. Ganduglia-Pirovano and M. Scheffler, Phys. Rev. B {\bf 59}, 15533 (1999).

\bibitem{todorova02}
M. Todorova, W.X. Li, M.V. Ganduglia-Pirovano, C. Stampfl, K. Reuter, and M. 
Scheffler, Phys. Rev. Lett. {\bf 89}, 096103 (2002).

\bibitem{li02}
W.X. Li, C. Stampfl and M. Scheffler, 
Phys. Rev. B {\bf 65}, 075407 (2002).

\bibitem{norskov90}
J.K. N{\o}rskov, Rep. Prog. {\bf 53}, 1253 (1990).

\bibitem{hammer01}
B. Hammer and J.K. N{\o}rskov, Adv. Catal. {\bf 45}, 71 (2001).

\bibitem{reuter02a}
K. Reuter, M.V. Ganduglia-Pirovano, C. Stampfl, and M. Scheffler,
Phys. Rev. B {\bf 65}, 165403 (2002).

\bibitem{li03}
W.X. Li, C. Stampfl and M. Scheffler,
Phys. Rev. B {\bf 67}, 045408 (2003).

\bibitem{ganduglia02}
M.V. Ganduglia-Pirovano, K. Reuter and M. Scheffler, Phys. Rev. B {\bf 65}, 
245426 (2002).
	
\bibitem{todorova04}
M. Todorova, K. Reuter and M. Scheffler, Phys. Rev. B ({\em in preparation}).

\bibitem{gibson99}
K.D. Gibson, M. Viste, E.C. Sanchez, and S.J. Sibener,
J. Chem. Phys. {\bf 110}, 2757 (1999). 

\bibitem{stampfl-PRL-1996}
C. Stampfl, S. Schwegmann, H. Over, M. Scheffler, and G. Ertl,
Phys. Rev. Lett. {\bf 77}, 3371 (1996).

\bibitem{wider98}
J. Wider, T. Greber, E. Wetli, T.J. Kreutz, P. Schwaller, and J. Osterwalder, Surf. Sci. {\bf 417}, 301 (1998).

\bibitem{carlisle00a}
C.I. Carlisle, T. Fujimoto, W.S. Sim, and D.A. King, 
Surf. Sci. {\bf 470}, 15 (2000); and references therein.

\bibitem{boettcher97}
A. B\"ottcher, H. Niehus, S. Schwegmann, H. Over, and G. Ertl,
J. Phys. Chem. B {\bf 101}, 11185 (1997).

\bibitem{over00}
H. Over, Y.D. Kim, A.P. Seitsonen, S. Wendt, E. Lundgren, M. Schmid,
P. Varga, A. Morgante, and G. Ertl, Science {\bf 287}, 1474 (2000).

\bibitem{gustafson04}
J. Gustafson, A. Mikkelsen, M. Borg, E. Lundgren, L. K\"ohler, G. Kresse,
M. Schmid, P. Varga, J. Yuhara, X. Torrelles, C. Quiros, and J.N. Andersen,
Phys. Rev. Lett. {\bf 92}, 126102 (2004).

\bibitem{lundgren02}
E. Lundgren, G. Kresse, C. Klein, M. Borg, J.N. Andersen, M. De Santis, Y. 
Gauthier, C. Konvicka, M. Schmid, and P. Varga, 
Phys. Rev. Lett. {\bf 88}, 246103 (2002).

\bibitem{carlisle00b}
C.I. Carlisle, D.A. King, M.L. Boucquet, J. Cerd{\'a}, and P. Sautet, 
Phys. Rev. Lett. {\bf 84}, 3899 (2000).

\bibitem{zheng00}
G. Zheng and E.I. Altman, Surf. Sci. {\bf 462}, 151 (2000).

\bibitem{reuter02b}
K. Reuter, C. Stampfl, M.V. Ganduglia-Pirovano, and M. Scheffler, 
Chem. Phys. Lett. {\bf 352}, 311 (2002).

\bibitem{fromhold76}
A.T. Fromhold, {\em Theory of Metal Oxidation}, Vols. I \& II,
North-Holland, Amsterdam (1976).

\bibitem{roosendahl00}
S.J. Roosendahl, A.M. Vredenberg, and F.H.P.M. Habraken,
Phys. Rev. Lett. {\bf 84}, 3366 (2000).

\bibitem{michaelides03}
A. Michaelides, M.-L. Bocquet, P. Sautet, A. Alavi, and D.A. King,
Chem. Phys. Lett. {\bf 367}, 344 (2003).

\bibitem{reuter04b}
K. Reuter and M. Scheffler, Appl. Phys. A {\bf 78}, 793 (2004).

\bibitem{xie99}
J. Xie, S. de Gironcoli, S. Baroni, and M. Scheffler, 
Phys. Rev. B {\bf 59}, 970 (1999).

\bibitem{reuter02c}
K. Reuter and M. Scheffler, Phys. Rev. B {\bf 65}, 035406 (2002).

\bibitem{reuter03a}
K. Reuter and M. Scheffler, Phys. Rev. Lett. {\bf 90}, 046103 (2003).

\bibitem{reuter03b}
K. Reuter and M. Scheffler, Phys. Rev. B {\bf 68}, 045407 (2003).

\bibitem{hendriksen04}
B.L.M. Hendriksen, S.C. Bobaru and J.W.M. Frenken, Surf. Sci. {\bf 552}, 229 (2004); B.L.M. Hendriksen, {\em Model Catalysts in Action: High-Pressure Scanning Tunneling Microscopy}, Ph.D. thesis, Universiteit Leiden (2003).

\bibitem{rogal04}
J. Rogal, K. Reuter and M. Scheffler, ({\em in preparation}).

\bibitem{todorova03}
M. Todorova, E. Lundgren, V. Blum, A. Mikkelsen, S. Gray, J. Gustafson, M. Borg,
J. Rogal, K. Reuter, J.N. Andersen, and M. Scheffler, Surf. Sci. {\bf 541}, 101 (2003).

\bibitem{rogal04b}
J. Rogal, K. Reuter and M. Scheffler, Phys. Rev. B {\bf 69}, 075421 (2004).

\bibitem{lundgren04}
E. Lundgren, J. Gustafson, A. Mikkelsen, J.N. Andersen, A. Stierle, H. Dosch,
M. Todorova, J. Rogal, K. Reuter, and M. Scheffler, Phys. Rev. Lett. {\bf 92}, 046101 (2004).

\bibitem{bondzie96}
V.A. Bondzie, P. Kleban and D.J. Dwyer, Surf. Sci. {\bf 347}, 319 (1996).

\bibitem{hendriksen03a}
B.L.M. Hendriksen and J.W.M. Frenken, Phys. Rev. Lett. {\bf 89}, 046101 (2002).

\bibitem{over04}
H. Over and M. Muhler, Prog. Surf. Sci. {\bf 72}, 3 (2003).

\bibitem{reuter04c}
K. Reuter, D. Frenkel and M. Scheffler, Phys. Rev. Lett. {\bf 93}, 116105 (2004).

\bibitem{landau02}
D.P. Landau and K. Binder, {\em A guide to Monte Carlo simulations in statistical physics}, Cambridge University Press, Cambridge (2002).

\bibitem{kang95}
H.C. Kang and W.H. Weinberg, Chem. Rev. {\bf 95}, 667 (1995).

\bibitem{cant78}
N.W. Cant, P.C. Hicks and B.S. Lennon, J. Catal. {\bf 54}, 372 (1978).

\bibitem{peden86}
C.H.F. Peden and D.W. Goodman, J. Phys. Chem. {\bf 90}, 1360 (1986).

\bibitem{fan01}
C.Y. Fan, J. Wang, K. Jacobi, and G. Ertl, 
J. Chem. Phys. {\bf 114}, 10058 (2001).

\bibitem{madey75}
T.E. Madey, H.A. Engelhardt and D. Menzel, Surf. Sci. {\bf 48}, 304 (1975).

\bibitem{stampfl02} 
C. Stampfl, M.V. Ganduglia-Pirovano, K. Reuter, and M. Scheffler, 
Surf. Sci. {\bf 500}, 368 (2002).

\bibitem{ertl97}
{\em Handbook on Heterogeneous Catalysis}, G. Ertl, H. Kn\"ozinger
and J. Weitkamp (Eds.), Wiley, New York (1997).

\bibitem{kim01b}
Y.D. Kim, H. Over, G. Krabbes and G. Ertl, 
Topics in Catalysis {\bf 14}, 95 (2001).

\bibitem{li03b} 
W.X. Li, C. Stampfl and M. Scheffler, Phys. Rev. Lett. {\bf 90}, 256102 (2003).

\bibitem{li03c}
W.X. Li, C. Stampfl and M. Scheffler, Phys. Rev. B {\bf 68}, 165412 (2003). 

\bibitem{oh83}
S.H. Oh and J.E. Carpenter, J. Catal. {\bf 80}, 472 (1983).

\bibitem{kellog85}
G.L. Kellog, Phys. Rev. Lett. {\bf 54}, 82 (1985); 
Surf. Sci. {\bf 171}, 359 (1986).

\bibitem{peden88}
C.H.F. Peden, D.W. Goodman, D.S. Blair, P.J. Berlowitz, G.B. Fisher, and
S.H. Oh, J. Phys. Chem. {\bf 92}, 1563 (1988).

\end{references}
\end{document}